\documentclass[twocolumn,aps,prl,superscriptaddress,numerical,reprint,letterpaper]{revtex4}
\usepackage{amstext}
\usepackage{xcolor}
\usepackage{amssymb}
\usepackage{graphicx}
\usepackage{soul}
\usepackage[english]{babel}
\usepackage[utf8]{inputenc}
\usepackage{fancyhdr}
\usepackage{amsmath}

\makeatletter

\providecommand{\tabularnewline}{\\}

\@ifundefined{textcolor}{}
{%
 \definecolor{BLACK}{gray}{0}
 \definecolor{WHITE}{gray}{1}
 \definecolor{RED}{rgb}{1,0,0}
 \definecolor{GREEN}{rgb}{0,1,0}
 \definecolor{BLUE}{rgb}{0,0,1}
 \definecolor{CYAN}{cmyk}{1,0,0,0}
 \definecolor{MAGENTA}{cmyk}{0,1,0,0}
 \definecolor{YELLOW}{cmyk}{0,0,1,0}
}
\makeatother

\begin{document}
\title{Theory of Singlet Fission in Polyenes, Acene Crystals and Covalently Linked Acene Dimers}

\author{Karan Aryanpour}

\affiliation
{Department of Physics, University of Arizona, Tucson, Arizona 85721, United States}
\author{Alok Shukla}
\affiliation
{Department of Physics, Indian Institute of Technology, Powai, Mumbai 400076, India}
\author{Sumit Mazumdar}
\affiliation
{Department of Physics, University of Arizona, Tucson, Arizona 85721, United States}
\affiliation
{College of Optical Sciences, University of Arizona, Tucson, Arizona 85721, United States}
\date{\today}
\begin{abstract}
We report quadruple configuration interaction calculations within the extended Pariser$-$Parr$-$Pople Hamiltonian on the excited states of aggregates of polyenes, crystalline acenes, and covalently linked dimers of acene molecules. We determine the precise energy orderings and analyze the cluster wave functions in order to arrive at a comprehensive physical understanding of singlet fission in these diverse families of materials. Our computational approach allows us to retain a very large number of basis states and thereby obtain the correct relative energy orderings of one electron$-$one hole Frenkel and charge-transfer excitons versus intra- and intermolecular two electron$-$two hole triplet$-$triplet excited states. We show that from the energy orderings it is possible to understand the occurrence of singlet fission in polyene and acene crystals, as well as its near total absence in the covalently linked acene dimers. As in the acene crystals, singlet fission in the polyenes is a multichromophoric phenomenon, with the well-known 2$^1$A$_g^-$ playing no direct role. Intermolecular charge-transfer is essential for singlet fission in both acenes and polyenes, but because of subtle differences in the natures and orderings of the aggregate excited states, the mechanisms of singlet fission are slightly different in the two classes. We are thus able to give qualitative physical reasoning for the slower singlet fission in the polyenes, relative to that in crystalline pentacene. Our work also gives new insight into the complex exciton dynamics in tetracene crystals, which has been difficult to understand theoretically. Our large-scale many-body calculations provide us with the ability to understand the qualitative differences in the singlet fission yields and rates between different classes of $\pi$-conjugated materials.
\end{abstract}
\maketitle
\section*{$\blacksquare$ INTRODUCTION}
\label{sec:intro} 
\begin{figure}
\includegraphics[width=3.0in]{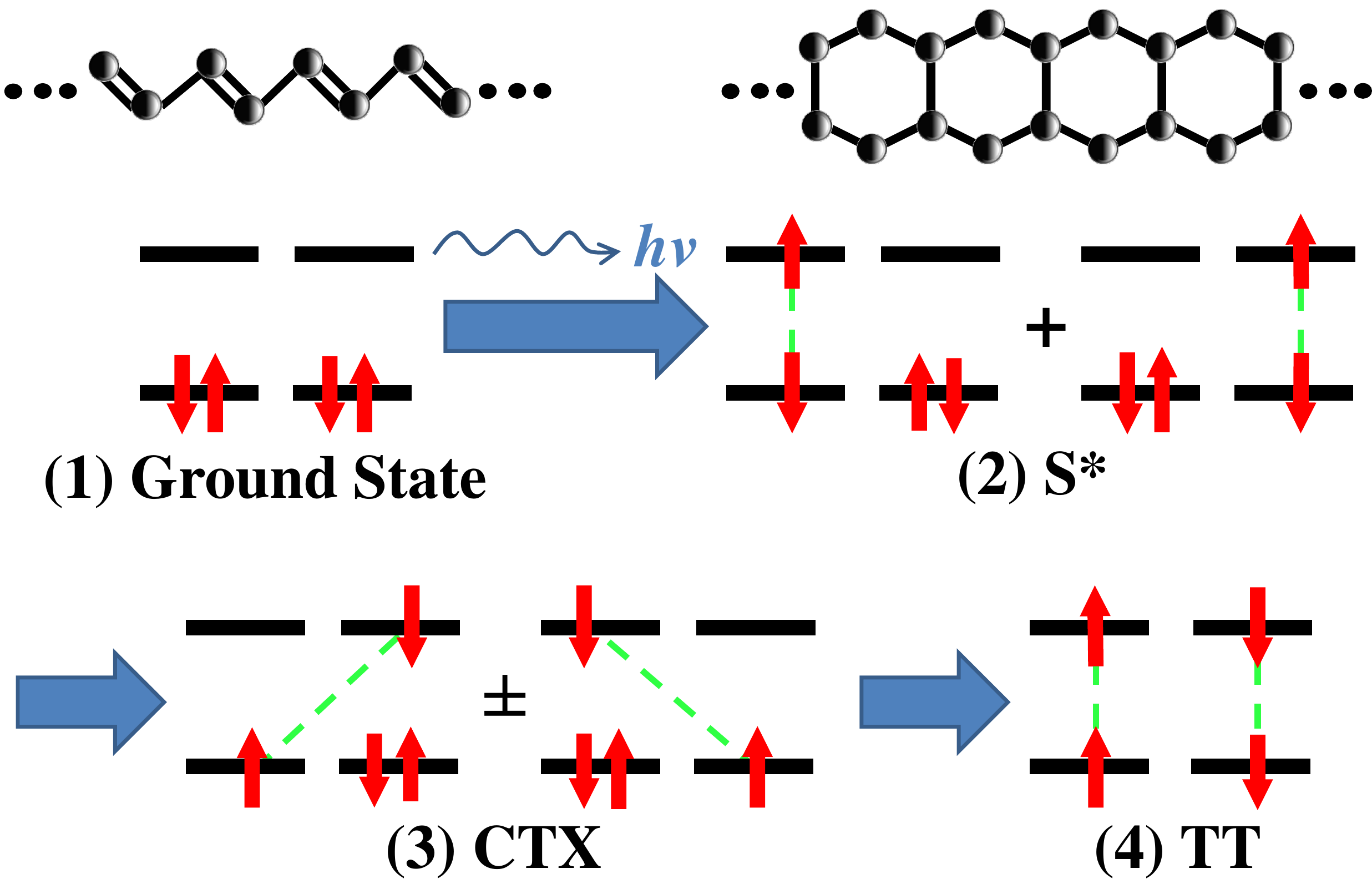}
\\ \center{Table of Contents}
\label{TOC}
\end{figure}
\par Singlet fission (SF) in organic materials, the process by which an optically excited spin$-$singlet intramolecular state S* dissociates into two lower-energy spin triplet excitons T, has been the focus of extensive investigation over the past few years. \cite{Smith10a,Smith13a} Provided specific energetic considerations are met in an organic donor$-$acceptor heterostructure, \cite{Aryanpour13a} each long-lived triplet exciton generated through SF in the donor molecule can in principle undergo charge dissociation at the donor$-$acceptor interface, thereby doubling the number of charge carriers compared to that obtained from the dissociation of the singlet exciton. SF-driven strong enhancement of performance has been found in pentacene$-$C$_{60}$ solar cells, \cite{Lee09a,Congreve13a} and the process can clearly have a significant impact on the search for organic solar cells with even higher quantum efficiency. \cite{Jadhav12a,Ehrler12a} The fundamental requirement for efficient SF, based on energetic considerations, is $E($S*$)\geqslant2\times E($T$_{1})$, where $E($S*$)$ and $E($T$_{1})$ are the energies of the singlet optical exciton and the lowest triplet exciton, respectively. Among various classes of organic materials fulfilling this requirement, \cite{Smith10a,Smith13a} aggregates of tetracene and pentacene, \cite{Jundt95a,Lee09a,Congreve13a,Thorsmolle09a,Burdett10a,Rao10a,Chan11a,Wilson11a,Chan12a,Roberts12a,Burdett11a,Tayebjee13a,Wilson13a,Burdett12a} acene derivatives, \cite{Walker13a,Ramanan12a,Ma12a,Yost14a} carotenoids and polyene crystals, \cite{Gradinaru01a,Papagiannakis02a,Wang10a,Wang12a,Dillon13a} and 1,3-diphenylisobenzofuran \cite{Johnson10a} exhibit SF with T yield ranging from 100\% to 200\%. There also exist tetracene and pentacene derivatives which {\it do not} undergo SF. \cite{Muller06a,Muller07a} A unified theoretical understanding of SF in the various systems is clearly desirable for the synthesis of new materials exhibiting SF and with potential use in solar cells. This is the goal of the current theoretical research.
\par Existing theories of SF \cite{Kuhlman10a,Zimmerman10a,Zimmerman11a,Havenith12a,Berkelbach13b,Chan13a,Beljonne13a,Zimmerman13a,Kolomeisky13a,Feng13a,Zeng14a,Parker14a,Yost14a} have almost universally focused on acenes (on pentacene in most cases), where it is generally agreed upon that SF is a multichromophoric process, and the triplets are generated from a triplet$-$triplet (hereafter $^1$(TT)) state with two triplet excitons localized on different chromophores and with overall spin angular momentum of zero. The $^1$(TT) state is also referred to as a multiexciton (hereafter ME) in the literature, as it is predominantly a correlated two electron$-$two hole (2e$-$2h) excitation from the ground state. The single question that has been investigated most intensively in this context is whether or not charge-transfer (CT) excitons (hereafter CTX) mediate the S* to $^1$(TT) conversion. \cite{Kuhlman10a,Greyson10a,Zimmerman10a,Zimmerman11a,Zimmerman13a,Teichen12a,Beljonne13a,Chan13a,Berkelbach13b} Some investigators have claimed that the $^1$(TT) state is reached from S* in a single step, driven by Coulomb interactions consisting of two-particle intermolecular electron transfers. \cite{Kuhlman10a,Zimmerman10a,Zimmerman11a,Zimmerman13a} The origin of this idea lies in the especially rapid SF in pentacene, where SF occurs in $\sim$ 80 fs. \cite{Wilson11a} According to these authors, mediation by the CTX, being necessarily a two-step process (S* $\to$ CTX $\to$ $^1$(TT)), would make $^1$(TT) generation and the SF following it much slower. This viewpoint has been contradicted by the proponents of the CTX-mediated mechanism, \cite{Smith10a,Greyson10a,Teichen12a,Beljonne13a,Chan13a,Berkelbach13b} who believe that the only requirement for the two-step process to be ultrafast is that S* and $^1$(TT) are of nearly the same energy. Additional justification of the CTX-mediated process comes from the matrix element calculations of the Coulomb interaction involving two-particle hops, which have found this matrix element too small to explain femtosecond ME generation. \cite{Chan13a,Smith13a,Berkelbach13b,Parker14a,Zeng14a} 
\par Theories emphasizing CT also differ in details themselves. A number of them find the CTX significantly higher in energy than S*, \cite{Zimmerman10a,Berkelbach13b,Zimmerman11a,Zimmerman13a} in which case a superexchange like mechanism of $^1$(TT) generation from S* has been suggested. Yet other mechanisms have also been proposed. Time-resolved experiments have been interpreted as showing direct photoexcitation of a coherent superposition of S* and $^1$(TT). \cite{Chan11a} Yamagata \cite{Yamagata11a} and Beljonne et al., \cite{Beljonne13a} based on their fitting of the Davydov splittings in the absorption spectra of the acenes, claim that optical excitation here is to a state that is a quantum mechanical superposition of S* and CTX and that transition from this optically accessible state to the $^1$(TT) occurs in a virtual one-step process. These authors have however also found that for intermolecular separation {\it smaller} than the equilibrium distance (i.e., very large intermolecular hopping) direct optical excitation to a state that is a superposition of S*, CTX, and $^1$(TT) is possible. Zeng et al. \cite{Zeng14a} arrive at the same conclusion, but once again, only for artificially large intermolecular couplings. Finally, it has been suggested that SF occurs directly from S*, without going through any real or virtual intermediate state. \cite{Yost14a}
\par The above disagreements are largely due to the absence of understanding of the precise natures and relative energies of the electronic states relevant to SF. Equally importantly, there exist several additional questions beyond those that have been investigated so far. First, how do morphology and environment influence SF? It has been widely observed experimentally that structural arrangement of the chromophores often determines whether or not SF efficiency is high. \cite{Smith10a,Smith13a,Dillon13a,Muller07a} Second, what is the mechanism of SF in carotenoid aggregates, \cite{Gradinaru01a,Papagiannakis02a,Wang10a,Wang12a,Dillon13a} in which the well-known intramolecular 2$^1$A$_{g}^{-}$ state, which is a quantum-entangled single-molecule state of two triplets, \cite{Ramasesha84a,Tavan87a,Baeriswyl92a} occurs below the optical 1$^1$B$_u^+$ exciton? \cite{Hudson82a,Christensen08a} (Here the superscript $1$ indicates spin singlet nature of the eigenstate while the ``plus'' and ``minus'' superscripts refer to the charge-conjugation symmetry). Is SF here monomolecular, with the triplet excitons originating from the dissociation of the 2$^1$A$_{g}^{-}$, or is the mechanism of triplet generation the same as in acenes? In either case, why is SF in carotenoids significantly slower \cite{Gradinaru01a,Papagiannakis02a,Wang10a,Wang12a,Antognazza10a} than in crystalline pentacene? Third, as has been noted by several research groups, although $E($S*$)<2\times E($T$_{1})$ in tetracene, \cite{Tomkiewicz71a} a simple activated mechanism for SF appears to fail. \cite{Burdett11a,Burdett12a,Tayebjee13a,Wilson13a} What precisely is behind this complex exciton dynamics? Finally, beyond the simple acenes and their derivatives, SF has also been experimentally studied in the so-called covalently linked acene dimers. \cite{Muller06a,Muller07a} However, unlike acene crystals or even disordered films, \cite{Roberts12a} little to no SF occurs in the dimer molecules. \cite{Muller06a,Muller07a} There have been few, if any, theoretical discussions of SF in covalently linked acene dimers; \cite{Smith13a,Berkelbach13b} whether or not the low yield of triplets in them can be understood within a comprehensive theory of SF remains an open question.
\par Our goal in the present work is to arrive at a broad theoretical framework for SF that can give qualitative and, wherever possible, semiquantitative answers to the above questions. Unlike existing quantum chemical approaches to SF, we make no attempt to calculate transition rates between excited states. This is because estimates of transition rates depend heavily on the calculated relative {\it energies} of initial and final states, which include the one electron-one hole (1e$-$1h) optical exciton and the CTX, as well as the 2e$-$2h $^1$(TT) state and the 2$^1$A$_g^-$. Treating 1e$-$1h and 2e$-$2h excitations on equal footing remains difficult for molecules with more than 8$-$10 electrons within first principles approaches. Getting the correct energy ordering of 2e$-$2h excitations in general requires configuration interaction (CI) with up to quadruple excitations (quadruple CI, hereafter QCI) from the ground state. \cite{Tavan86a,Aryanpour14a,Schmidt12a} Over and above in the present case such calculations have to be performed for aggregates (at least dimers) of large molecules and lie outside the scope of standard quantum chemical techniques. Theoretical approaches to SF therefore mostly involve calculations based on a few excitations across a limited number of frontier molecular orbitals (MOs) of pairs of molecules. We will show that such approximations can lead to errors in the quantum mechanical descriptions of the multichromophore eigenstates relevant in SF. For our modeling of chromophore clusters we therefore take the opposite approach here and employ the semiempirical Pariser$-$Parr$-$Pople (PPP) $\pi$-electron Hamiltonian. \cite{Pariser53a,Pople53a} Our approach allows us to perform CI calculations retaining up to quadruple excitations across a very large number of MOs (including the complete set in some cases), thereby ensuring accurate cluster excited state energy orderings. If one now makes the reasonable assumptions that (i) efficient transitions can occur between energetically proximate states and (ii) for SF to occur the $^1$(TT) should be either the {\it lowest} excited state of the cluster or must be close in energy to an optical state, then a physical understanding of SF can be obtained from the excited state energy spectrum alone.
\par The PPP Hamiltonian has successfully addressed correlation effects in a variety of single chromophore systems \cite{Chandross97a,Chandross99a,Wang06a} and in recent years has also reproduced experimental features in multichromophore systems. \cite{Aryanpour13a,Aryanpour10a,Aryanpour11a} In the present case, we will for the first time give a precise explanation of why SF in polyenes is multichromophoric, and the 2$^1$A$_g^-$ plays no role. A natural explanation of the much slower SF rate in polyenes compared to pentacene also emerges from our work. New insights on SF and exciton dynamics in tetracene, which have remained considerably less understood than in pentacene, are obtained. Finally, we give a clear explanation of the very low SF yield in covalently linked acene dimers. It is only through high-order CI studies of the PPP model that such a unified picture is obtained.
\section*{$\blacksquare$ THEORETICAL MODEL, COMPUTATIONAL METHODS AND PARAMETERIZATION}
\label{sec:model} 
\par\noindent\textbf{Model Hamiltonian.} We report QCI calculations for clusters of chromophores based on the modified PPP Hamiltonian consisting of {\it intra} and {\it inter}chromophore terms
\begin{equation}
\label{PPP_Ham}
H_{\mathrm{PPP}}=H_{\mathrm{intra}}+H_{\mathrm{inter}}
\end{equation}
with
\begin{eqnarray}
\label{intra_Ham}
 H_{\mathrm{intra}}=\sum_{\mu\langle ij \rangle\sigma}t_{ij}^{\mu}
(\hat{c}_{\mu i\sigma}^{\dagger}\hat{c}_{\mu j\sigma}^{}+\hat{c}_{\mu j\sigma}^\dagger \hat{c}_{\mu i\sigma}^{}) + U\sum_{\mu i}\hat{n}_{\mu i\uparrow} \hat{n}_{\mu i\downarrow} \nonumber \\
 + \sum_{\mu,i<j} V_{ij} (\hat{n}_{\mu i}-1)(\hat{n}_{\mu j}-1)\hspace{1.5in}
\end{eqnarray}
and
\begin{eqnarray}
\label{inter_Ham}
 H_{\mathrm{inter}}=\sum_{\mu<\mu',ij,\sigma}t^{\perp}_{ij}
(\hat{c}_{\mu i\sigma}^\dagger \hat{c}_{\mu' j\sigma}^{}+\hat{c}_{\mu' j\sigma}^\dagger \hat{c}_{\mu i\sigma}^{})+ \nonumber \\\frac{1}{2}\sum_{\mu<\mu',ij} V_{ij}^{\perp}(\hat{n}_{\mu i}-1)(\hat{n}_{\mu' j}-1)\hspace{0.5in}
\end{eqnarray}
where $\hat{c}^{\dagger}_{\mu i\sigma}$ creates a $\pi$-electron of spin $\sigma$ on carbon (C) atom $i$ located in chromophore $\mu$; $\hat{n}_{\mu i\sigma} = \hat{c}^{\dagger}_{\mu i\sigma}\hat{c}_{\mu i\sigma}^{}$ is the number of electrons of spin $\sigma$ on atom $i$ within chromophore $\mu$, and $\hat{n}_{\mu i}=\sum_{\sigma} \hat{n}_{\mu i\sigma}$. The intramolecular one-electron hopping integrals $t_{ij}^{\mu}$ are between nearest-neighbor C atoms $i$ and $j$. We have, however, chosen long-range intermolecular hopping integrals $t^{\perp}_{ij}$ between C atoms $i$ and $j$ (see below). $U$ is the Hubbard repulsion between two electrons with opposite spins occupying the same atomic $p_z$ orbital; $V_{ij}$ is the long-range intersite Coulomb interaction between two electrons on a single chromophore; and $V_{ij}^{\perp}$ is the corresponding interaction for different chromophores.
\begin{figure*}
\begin{flushleft}
{\footnotesize \bf Scheme 1: Nomenclature for the 1e$-$1h and 2e$-$2h Excitations that Dominate Eigenstates Energetically Proximate to the Optical Exciton in a Two-Chromophore (C1 and C2) System: (a) EXC, (b) P$^+$P$^-$ (c) COV and (d) (TT)$_n$$^a$}
\end{flushleft}
\includegraphics[width=4.5in]{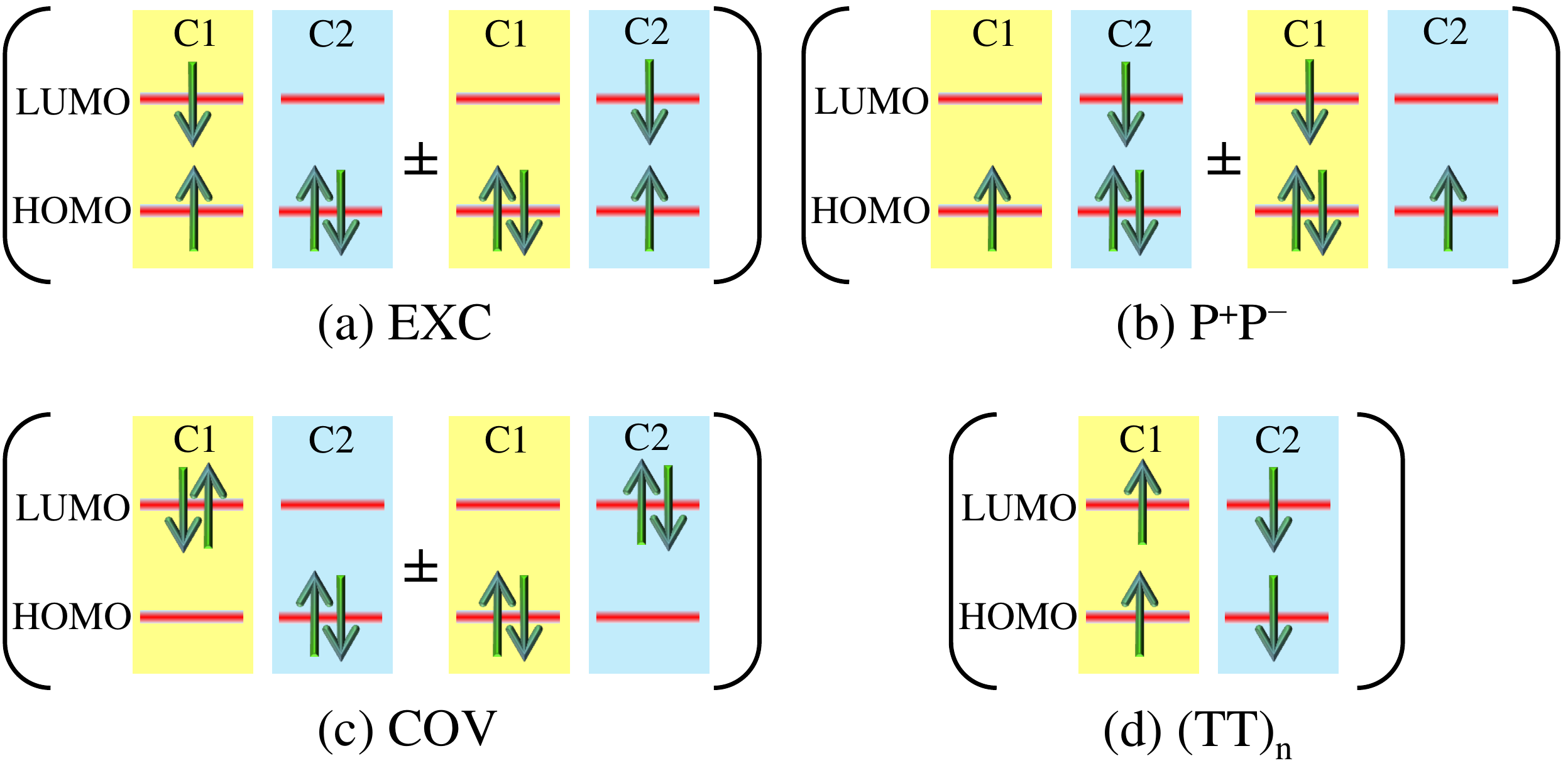}  
\label{s1} 
\begin{flushleft}
{\footnotesize $^a$Excitations across the HOMO and LUMO only are shown, while actual excited states may have excitations involving much lower bonding and higher antibonding MOs. Covalent excited states such as the 2$^1$A$_g^-$ have contributions from configurations of the type (c) as well as EXC configurations.}
\end{flushleft}
\end{figure*}
\par\noindent\textbf{Computational Approach.} We use a basis of {\it localized} Hartree$-$Fock (HF) MOs of the individual molecular units in our calculations. \cite{Aryanpour13a,Wang08a,Aryanpour10a,Aryanpour11a} This allows a clear distinction between excitations localized on a single molecule versus those that are delocalized over multiple molecules, as well as quantum mechanical superposition of these extreme configurations. For the very largest systems we are forced to ignore some of the outermost HF MOs in our QCI calculations. The number of MOs we have retained in all cases (cf. Supporting Information) is however significantly larger than what are retained within existing calculations. \cite{Zimmerman10a,Greyson10a,Zimmerman11a,Teichen12a,Havenith12a,Zimmerman13a,Berkelbach13a,Berkelbach13b,Zeng14a,Parker14a} For precise classification of all excited states proximate to the optical exciton in a multichromophore system, the QCI wave function of the $n$th excited state is expanded in terms of single, double, triple, and quadruple e$-$h excitation configurations
\begin{eqnarray}
\big|\Psi_{\mathrm{QCI}}^{(n)}\big>=\sum_{i}\xi_{i}^{(n,1)}\big|\mathrm{1e-1h}\big>_i+\sum_{i}\xi_{j}^{(n,2)}\big|\mathrm{2e-2h}\big>_j\nonumber \\+
\sum_{k}\xi_{k}^{(n,3)}\big|\mathrm{3e-3h}\big>_k+\sum_{l}\xi_{l}^{(n,4)}\big|\mathrm{4e-4h}\big>_l\hspace{0.5in}
\label{qci-wfn}
\end{eqnarray}
where within each class the $|\xi_{i}|$'s are sorted in a descending order. For each eigenstate we now calculate the wave function \textit{character density} $\rho=\sum_{|\xi_{i}|\ge0.1}|\xi_{i}|^{2}$, corresponding to each class of excitations $i$ (thus the calculated $\rho$ for a given eigenstate allows us to classify it as predominantly 1e$-$1h, predominantly 2e$-$2h, ... etc.). In what follows we ignore configurations with $|\xi_{i}|<0.1$ and also 3e$-$3h and 4e$-$4h excitations (3e$-$3h and 4e$-$4h excitations in any event have $|\xi_{i}|<0.1$). 
\par On the basis of such detailed analyses of the wave functions we are able to further classify \cite{Aryanpour10a,Aryanpour11a,Wang08a} eigenstates dominated by 1e$-$1h excitations as (i) Frenkel exciton (hereafter EXC), if the excited electron and hole are localized on a monomer, (ii) polaron pair (hereafter P$^+$P$^-$) if the electron and hole are completely charge separated and lie on different molecules, and (iii) CTX, if the wave function is a superposition of EXC and P$^+$P$^-$. Similarly, eigenstates dominated by 2e$-$2h excitations can have both excitations on the single monomer or on different monomers. In the energy region of interest (near and below the optical exciton), 2e$-$2h excitations localized on a monomer are the predominant constituents of the 2$^1$A$_g^-$ as well as of optically dark $^1$B$_u^-$ states (with additional contributions coming from configurations that are EXC). \cite{Ramasesha84a,Tavan87a,Baeriswyl92a} These excited states are quantum-entangled triplet$-$triplet states, where both triplet excitations occupy the same molecule. In the language of valence bond theory, these spin singlet excited states are \textit{covalent}, \cite{Ramasesha84a,Tavan87a} as opposed to the \textit{ionic} EXC states (2e$-$2h states can also be singlet$-$singlet; in this case they are doubly ionic, with two pairs of C$^+$ and C$^-$ ions; these have energies nearly twice that of the optical exciton \cite{Chandross99a} and are not of interest here). We will label such monomer 2e$-$2h $^1$A$_g^-$ and $^1$B$_u^-$ excitations collectively as COV. Finally, there exist also eigenstates dominated by 2e$-$2h excitations in which two 1e$-$1h excitations occur on different monomers. These are overall spin singlet excitations consisting of two triplets on two monomers, which we will label as (TT)$_n$, where $n$ is a measure of the separation between the triplet excitations ($n=1$ corresponds to nearest-neighbor triplets, $n=2$ as next nearest neighbor, ..., etc.). In Scheme~\ref{s1}a$-$d we have given the schematics of EXC, P$^+$P$^-$, COV, and  (TT)$_n$ excitations, respectively, for a dimer of two chromophores C1 and C2, where, however, we have included only the highest occupied and the lowest unoccupied MOs (HOMO and LUMO, respectively) in describing them. The actual wave functions can have contributions from excitations across much larger one-electron gaps. Note that the (TT)$_n$ configuration is a normalized superposition of three fundamental configurations, and the formal expression in the limited HOMO$-$LUMO basis of Scheme~\ref{s1} is
\begin{eqnarray}
(TT)_{n}=\frac{1}{\sqrt{3}}\Big[\sum_{\sigma}a_{1L\sigma}^\dagger a_{1H,-\sigma}^{}a_{2L,-\sigma}^\dagger a_{2H\sigma}^{} + \nonumber \\
\frac{1}{2}(a_{1L\downarrow}^\dagger a_{1H\downarrow}^{}-a_{1L\uparrow}^\dagger a_{1H\uparrow}^{})(a_{2L\downarrow}^\dagger a_{2H\downarrow}^{}-a_{2L\uparrow}^\dagger a_{2H\uparrow}^{})\Big]|G\rangle
\label{TT-wfn}
\end{eqnarray}
where $a_{iH\sigma}^\dagger$ ($a_{iL\sigma}^\dagger$) creates an electron in the HOMO (LUMO) of molecule $i=1,2$ and $|G\rangle$ is the HF ground state. There are three distinct terms (since $\sigma=\uparrow$, $\downarrow$) which correspond to ($S_z^1$, $S_z^2$) = ($+1,-1$); ($-1,+1$); and ($0,0$) excitations on the individual units, where $S_z^i$ is the $z$ component of the spin on molecule $i$. Using the same limited HOMO$-$LUMO basis, the other three classes of excitations in Scheme~\ref{s1} are written as
\begin{subequations}
\begin{align}
EXC=\frac{1}{2}\sum_{\sigma}\big(a_{1L\sigma}^\dagger a_{1H\sigma}^{} \pm a_{2L\sigma}^\dagger a_{2H\sigma}^{}\big)|G\rangle \hspace{0.5in} \\
P^+P^-=\frac{1}{2}\sum_{\sigma}\big(a_{2L\sigma}^\dagger a_{1H\sigma}^{} \pm a_{1L\sigma}^\dagger a_{2H\sigma}^{}\big)|G\rangle \hspace{0.5in} \\
COV=\frac{1}{\sqrt{2}}\big(a_{1L\uparrow}^\dagger a_{1L\downarrow}^\dagger a_{1H\downarrow}^{} a_{1H\uparrow}^{} \pm a_{2L\uparrow}^\dagger a_{2L\downarrow}^\dagger a_{2H\downarrow}^{} a_{2H\uparrow}^{}\big)|G\rangle
\end{align}
\label{Other-wfn}
\end{subequations}
Only the ``plus'' superposition in eq~\ref{Other-wfn}a is dipole allowed; the ``minus'' superposition is an optically ``dark'' exciton. CTX states are superpositions of EXC and P$^+$P$^-$ with varying relative weights.
\par For further clarifications of our labeling scheme we note the following. First, multiple-chromophore systems are expected to have multiple excitonic states that are dipole-allowed, which is the origin of the experimentally observed Davydov splitting in these systems. Second, however, not all EXC states are optically allowed. Particularly when the two molecules in a dimer cluster are related by any symmetry operation, the odd superposition of the excitations on two molecules that are otherwise equivalent will be optically dark.
\par {\bf Parameters.} Our parametrizations of $H_{intra}$ are as follows. The intramolecular hopping integrals are $-2.4$ eV for the phenyl C$-$C bonds unless otherwise stated and $-2.2$ ($-2.6$) eV for the single (double) bonds. These values are accepted as standard within correlated-electron models \cite{Ramasesha84a,Tavan87a,Baeriswyl92a} (noninteracting H\"uckel models often use larger hopping integrals $-2.7$ to $-3.0$ eV in order to fit experimentally observed optical absorptions; such large values are neither appropriate nor necessary when Coulomb interactions are nonzero). The intramolecular intersite interactions are obtained from parametrization of the PPP model obtained previously, \cite{Chandross97a} $V_{ij}=U/\kappa\sqrt{1+0.6117 R_{ij}^2}$,  where $R_{ij}$ is the distance in $\mathring{\textrm{A}}$ between C atoms $i$ and $j$ and $\kappa$ is an effective dielectric constant. For the linear polyene calculations reported below, we use $U=8.0$ eV and $\kappa=2$. These parameters have been used extensively in the past, \cite{Chandross97a,Wang06a} including for linear polyenes, \cite{Aryanpour14a} in all cases with excellent fits to experimentally determined singlet and triplet excitation energies. The Coulomb interaction parameters that we chose for our calculations on the acenes are slightly smaller, in order to reproduce the experimentally obtained energies of the spin triplet excitons (see below).
\par For $H_{inter}$, we have chosen the same functional form for the intermolecular Coulomb interactions $V_{ij}^{\perp}$ as in $H_{intra}$. The value of the intermolecular screening parameter $\kappa^{\perp}$ is uncertain; we have taken this to be the same as the intramolecular screening parameter $\kappa$ for simplicity. The intermolecular hopping integrals between C atoms $i$ and $j$ are obtained by adjusting $\beta$ in the expression \cite{Aryanpour10a,Uryu04a}
\begin{equation}
\label{inter_hop}
t^{\perp}_{ij}=\beta~\mathrm{exp}[(h_{min}-R_{ij})/\delta](\frac{\hat{\bf n}_1\cdot\vec{\bf R}_{ij}}{R_{ij}})(\frac{\hat{\bf n}_2\cdot\vec{\bf R}_{ij}}{R_{ij}})
\end{equation}
with $h_{min}$ being the distance between the closest two C atoms located on two separate chromophores, $\beta$ the hopping integral between those two C atoms, and $\delta=0.045$ nm. Unit vectors $\hat{\bf n}_1$ and $\hat{\bf n}_2$ are normal to the surfaces of the chromophores 1 and 2 on atoms $i$ and $j$, respectively. For our calculations on polyenes, we set $\beta=-0.2$ eV, which has been previously used successfully for multiple-walled carbon nanotubes. \cite{Uryu04a} For acenes, we have performed calculations for the range of $0\leqslant|\beta|\leqslant0.2$ eV (cf. Supporting Information for details) to reproduce known experimental results. As mentioned above, and as we shall see from our calculations below, the parameters we have chosen can be further justified {\it a posteriori}, based on our ability to explain SF and associated trends in all the compounds we consider.
\begin{figure*}
\includegraphics[width=6.0in]{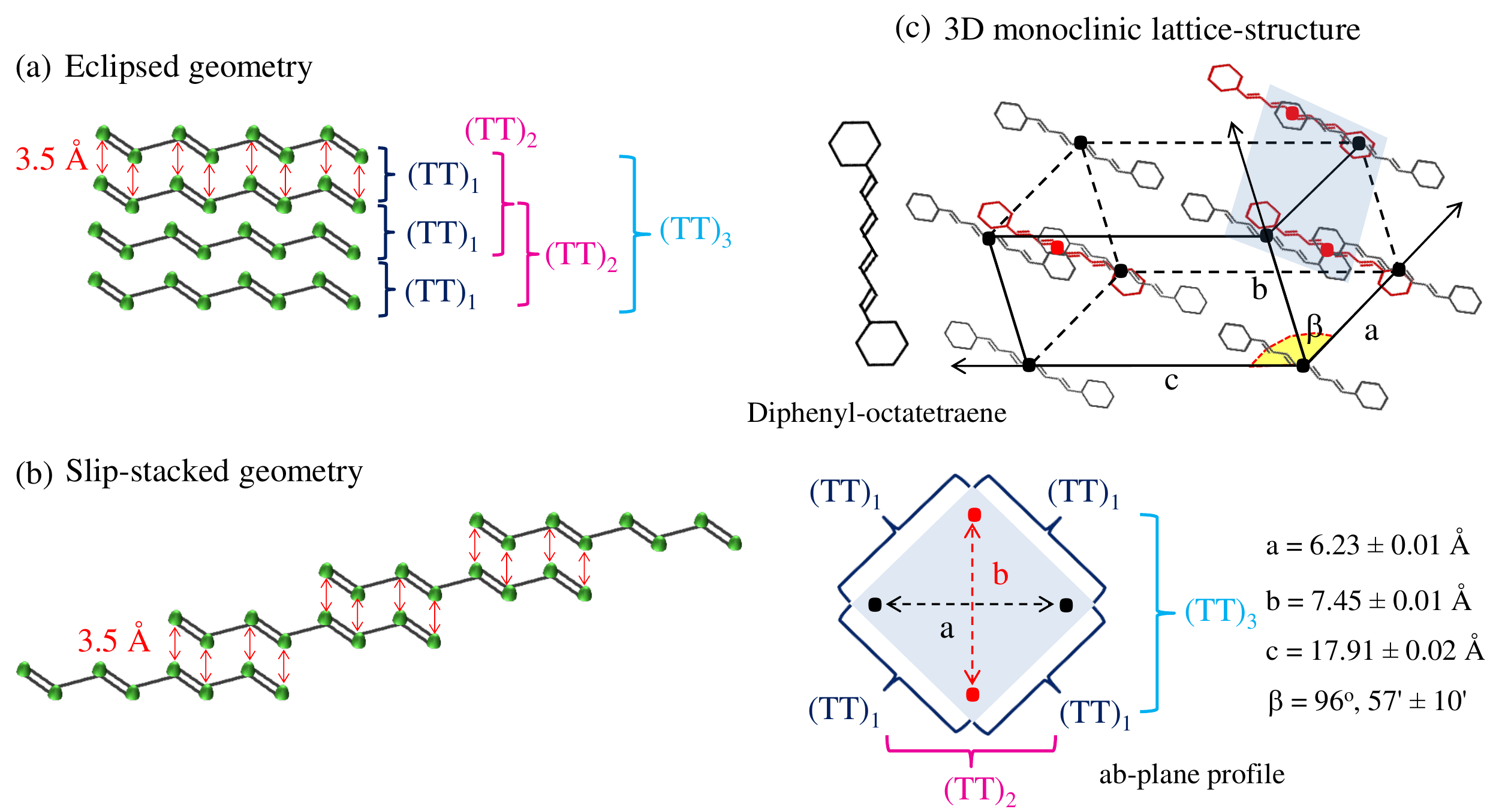} 
\caption{Polyene clusters considered in this work. (a) Eclipsed, in which the molecules lie in parallel planes, separated by $3.5~\mathring{\textrm{A}}$; each atom lies directly on top of another in the next plane, when viewed vertically. The nomenclature for the intermolecular triplet$-$triplet states (TT)$_{n}$ is defined in the figure. (b) Slip-stacked, generated by sliding each molecule in (a) from the second molecule onward by half the molecular length along the direction of the molecular axis. The (TT)$_{n}$ states are defined as in (a). An alternate zigzag slip-stacked geometry, in which each atom of the second neighbor molecule, but not the nearest neighbor, lies directly on top of the corresponding atom of the original molecule, gives very similar results as for this case. (c) The four molecules that constitute the nearest and next nearest neighbors of one another within the monoclinic crystal structure of carotenoid crystals. The molecules occupy the vertices of the gray parallelogram. The separations between the molecules, as well as the different (TT)$_n$ states, are indicated in the figure below. Lattice constants and angle $\beta$ have been taken from refs~\citenum{Drenth54a} and \citenum{Weiss97a}.}
\label{f1} 
\end{figure*}
\section*{$\blacksquare$ RESULTS AND ANALYSIS}
\label{sec:results}
\begin{table*}
\begin{tabular}{l}
{\footnotesize \bf Table 1: Calculated Energies in eV of the Optical Exciton, T$_{1}$, Lowest (TT)$_{n}$ and COV States of Clusters of Octatet-}\tabularnewline 
{\footnotesize \bf raene Molecules$^a$}\tabularnewline
\hline \hline
{\footnotesize chromophore number\hspace{1.0cm}optical exciton ($|\mu|$)\hspace{2.3cm}T$_{1}$ \hspace{2.0cm}(TT)$_{n}$ \hspace{1.85cm}$2^{1}$A$_{g}^{-}$ \hspace{1.9cm}$1^{1}$B$_{u}^{-}$}\tabularnewline
{\footnotesize 2 (eclipsed)\hspace{3.0cm}4.94 (2.35)\hspace{2.7cm}1.79\hspace{2.2cm}3.65\hspace{2.2cm}3.77\hspace{2.2cm}4.67}\tabularnewline
{\footnotesize\hspace{8.60cm}1.83\hspace{4.9cm}3.81\hspace{2.2cm}4.71}\tabularnewline
{\footnotesize 3 (eclipsed)\hspace{3.0cm}4.72 (1.88)\hspace{2.7cm}1.69\hspace{2.2cm}3.58\hspace{2.17cm}4.00\hspace{2.22cm}5.18}\tabularnewline
{\footnotesize\hspace{4.5cm}4.74 (2.46)\hspace{2.7cm}1.75\hspace{4.9cm}4.07\hspace{2.2cm}5.26}\tabularnewline
{\footnotesize\hspace{8.60cm}1.78\hspace{4.9cm}4.15\hspace{2.2cm}5.28}\tabularnewline
{\footnotesize 4 (eclipsed)\hspace{3.0cm}4.41 (2.91)\hspace{2.7cm}1.62\hspace{2.2cm}3.42\hspace{2.17cm}3.98\hspace{2.20cm}5.84}\tabularnewline
{\footnotesize\hspace{4.5cm}4.47 (2.39)\hspace{2.7cm}1.67\hspace{4.9cm}4.04\hspace{2.2cm}5.87}\tabularnewline
{\footnotesize\hspace{8.60cm}1.72\hspace{4.9cm}4.11\hspace{2.2cm}6.00}\tabularnewline
{\footnotesize\hspace{8.60cm}1.74\hspace{4.9cm}4.20\hspace{2.2cm}6.02}\tabularnewline
{\footnotesize 4 (slip-stacked)\hspace{2.5cm}4.19 (3.65)\hspace{2.7cm}1.70\hspace{2.2cm}3.57\hspace{2.19cm}4.16\hspace{2.18cm}5.94}\tabularnewline
{\footnotesize\hspace{4.5cm}4.20 (1.60)\hspace{2.7cm}1.71\hspace{4.9cm}4.19\hspace{2.2cm}5.97}\tabularnewline
{\footnotesize\hspace{8.6cm}1.73\hspace{4.9cm}4.22\hspace{2.2cm}6.01}\tabularnewline 
{\footnotesize\hspace{8.6cm}1.73\hspace{7.62cm}6.07}\tabularnewline
{\footnotesize 4 (monoclinic)\hspace{2.6cm}4.32 (3.88)\hspace{2.7cm}1.73\hspace{2.2cm}3.63\hspace{2.19cm}4.18\hspace{2.18cm}5.92}\tabularnewline
{\footnotesize\hspace{8.60cm}1.73\hspace{4.9cm}4.18\hspace{2.2cm}5.97}\tabularnewline
{\footnotesize\hspace{8.60cm}1.73\hspace{4.9cm}4.18\hspace{2.2cm}6.01}\tabularnewline
{\footnotesize\hspace{8.60cm}1.73\hspace{4.9cm}4.26\hspace{2.2cm}6.04}\tabularnewline
\hline \hline 
\end{tabular} \\
{\footnotesize $^a$The single-molecule energies are $E$(T$_{1})=1.65$ eV, $E(2^{1}$A$_{g}^{-})=3.42$ eV, $E(1^{1}$B$_{u}^{-})=4.28$ eV and $E(1^{1}$B$_{u}^{+})=4.50$ eV. The numbers in parentheses in the first column are the transition dipole couplings between the optically allowed states and the ground state, in $\textrm{e}\mathring{\textrm{A}}$. In the slip-stacked geometry there are no pure $g$ or $u$ states and the 2$^1$A$_g^-$-derived states are weakly dipole coupled to the ground state.}
\label{t1} 
\end{table*}
\par\noindent\textbf{Linear Polyenes.} In carotenoids, the linear polyene backbone of alternant single and double bonds accounts for many of the physical properties related to their excited states. \cite{Polivka04a} In what follows, the benzene rings at both ends of diphenylpolyene molecules have been omitted to reduce the number of C atoms in our molecular clusters to what is numerically tractable. QCI calculations were performed for clusters of up to four octatetraene and three decapentaene molecules. We show the results only for octatetraene, as the results for decapentaene were similar in all cases. We investigated three different cluster geometries: (a) eclipsed, in which the polyene molecules are symmetrically stacked on top of each other (Figure~\ref{f1}a), (b) slip-stacked, with molecules from the second layer onward shifted rigidly along the molecular axis (Figure~\ref{f1}b), and (c) monoclinic (Figure~\ref{f1}c). The eclipsed and slip-stacked geometries are unrealistic; our reason for adopting them is because they allow us to visualize the effects of aggregation in a systematic manner, by adding monomers one at a time. As seen below, calculations with these structures also help in explaining why efficient SF may not occur even when energetic considerations are satisfied. Experimentally, carotenoid crystals have either the monoclinic or the orthorhombic structure, \cite{Drenth54a,Weiss97a} and the former is associated with efficient SF. \cite{Dillon13a}
\par
\begin{figure*}
\includegraphics[width=7in]{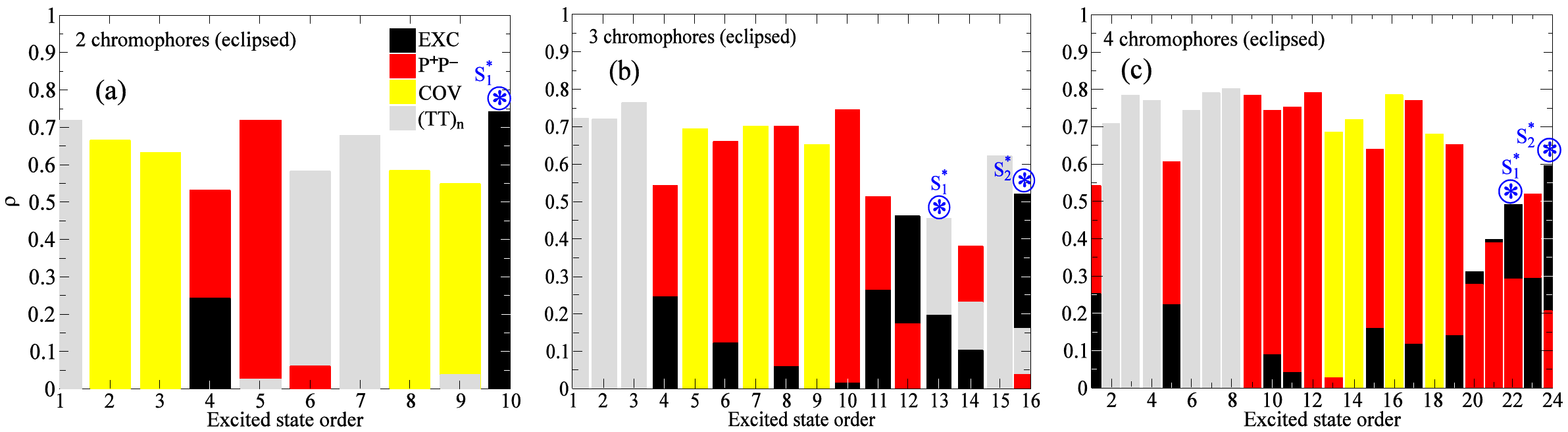} 
\protect\caption{Wave function character density $\rho$ (see text) for the excited states below the optical exciton in the eclipsed polyene cluster, for the (a) dimer, (b) trimer, and (c) tetramer, respectively. The different color schemes define the EXC (Scheme~\ref{s1}a), P$^+$P$^-$ (Scheme~\ref{s1}b), COV (Scheme~\ref{s1}c) and (TT)$_{n}$ (Scheme~\ref{s1}d) contributions. Circled asterisks identify optical excitons with significant transition dipole couplings to the ground state (cf. Table~\ref{t1}). The lowest state in the tetramer is a CTX and not (TT)$_{n}$.}
\label{f2} 
\end{figure*}

 {\it Eclipsed Geometry.} Our computational results for this case are summarized in Table~\ref{t1} and Figure~\ref{f2}. Table~\ref{t1} lists the energies of different kinds of excited states, excluding the P$^+$P$^-$ and CTX excitations. The excited states are identified from comparisons of the energies of the single-molecule and multiple-molecule eigenstates. These identifications are further confirmed from the analyses of the multiple-molecule wave functions, as done in Figure~\ref{f2}. In every case in Figure~\ref{f2}, $\rho$ gives the normalized weight of particular excitation components within the wave function of excited states indicated in the figure (thus, for example, the contribution by the (TT)$_{n}$ excitation components to the lowest excited state of two molecules of octatetraene in Figure~\ref{f2}a is $0.72$. In reality the (TT)$_{n}$ contribution would have been even larger had we not ignored 3e$-$3h and 4e$-$4h excitations in our computation of $\rho$. Similarly the next two eigenstates in the two-chromophore cluster in Figure~\ref{f2}a are derived from the single molecule 2$^1$A$_g^-$ state and so on). States 4, 5, and 6 in Figure~\ref{f2}a are seen to be of mixed character. Thus, state 4 is CTX, with P$^+$P$^-$ contribution of $\sim$ 0.5 and EXC contribution of $\sim$ 0.25. Similarly state 6 is $\sim$ 0.55 (TT)$_n$ and $\sim$ 0.05 P$^+$P$^-$ and so on.
\par A number of features and trends are noteworthy regarding the results presented in Table~\ref{t1} and Figure~\ref{f2}. First, there is little to no mixing between eigenstates that are predominantly 1e$-$1h and those that are predominantly 2e$-$2h when the number of molecules is even. The absence of coherent superposition of EXC and (TT)$_n$, seen with even numbers of chromophores, persists for other geometries and other chromophores (see below). Of course such a superposition can be obtained for $t_{ij}^{\perp}$ much larger than assumed in our work, but as we will see our parametrized $t_{ij}^{\perp}$ with $\beta=-0.2$ eV already gives Davydov splittings larger than that observed experimentally and hence gives the upper limit to realistic $t_{ij}^{\perp}$. The optical excitons are easily identified in all cases from their nonzero transition dipole couplings with the multichromophoric ground state, as is indicated in Table~\ref{t1}. As seen in Figure~\ref{f2}, among eigenstates with significant EXC character, the highest energy states are allowed optically, indicating the H-aggregate character of the eclipsed geometry. The energy of the optical exciton decreases with increasing number of molecules, a well-known effect that is in agreement with the experimentally observed redshift of the optical state in the solid. Our calculations also reproduce qualitatively the Davydov splitting between optical excitons expected in clusters. This is the energy difference between S$_1^{\textstyle{*}}$ and S$_2^{\textstyle{*}}$ in the figures.
\par Our criterion here for determining that a given morphology is conducive to SF is that the lowest excited state necessarily be (TT)$_n$ (see however below for discussion of the tetracene excitation spectrum). Of particular interest in the present context therefore are the relative energies of the (TT)$_n$, COV, and the CTX eigenstates. We have found that the multichromophoric (TT)$_n$ states not only include the lowest triplets T$_1$ but also can be composed of the next higher triplet T$_2$ (this is why the number of (TT)$_1$ states in Figure~\ref{f2}a exceeds $2$). The COV states are in all cases higher in energy than the (TT)$_n$. These are derived from both the 2$^1$A$_g^-$ and the 1$^1$B$_u^-$ in Figure~\ref{f2}a and from the 2$^1$A$_g^-$ alone in Figure~\ref{f2}b, c. As seen in Table~\ref{t1}, interchain interactions {\it increase} the energies of the COV states 2$^1$A$_g^-$ and 1$^1$B$_u^-$, while the shift in the energies of T$_1$ as well as (TT)$_n$ is very weak. The increase in energies of the COV states, relative to the lowest (TT)$_n$, has obvious significance for SF. We have therefore examined this very carefully, by performing QCI calculations for clusters of hexatrienes, for which for up to three chromophores all MOs could be retained in the calculations. Results identical to those shown in Table~\ref{t1} were found (cf. Supporting Information). Furthermore, with increasing number of chromophores, we have ascertained that the (TT)$_n$ states are superpositions of $n=1$, $2$ and $n=3$, indicating the tendency to give a (TT)$_{n}$ band in which the two triplets are not bound. While all of these suggest efficient SF in this geometry, as indicated in Figure~\ref{f2}, the optically dark CTX states progressively shift downward in energy with an increase in the number of chromophores and constitute the lowest-energy state for the four monomer case. The hypothetical eclipsed geometry is thus not conducive for SF, as the lowest excited state here is an excimer (the CTX is the lowest excited state of three eclipsed decapentaene chromophores in our calculations, indicating the role of chromophore size in addition to the geometry).
\par
\begin{figure}
\includegraphics[width=3in]{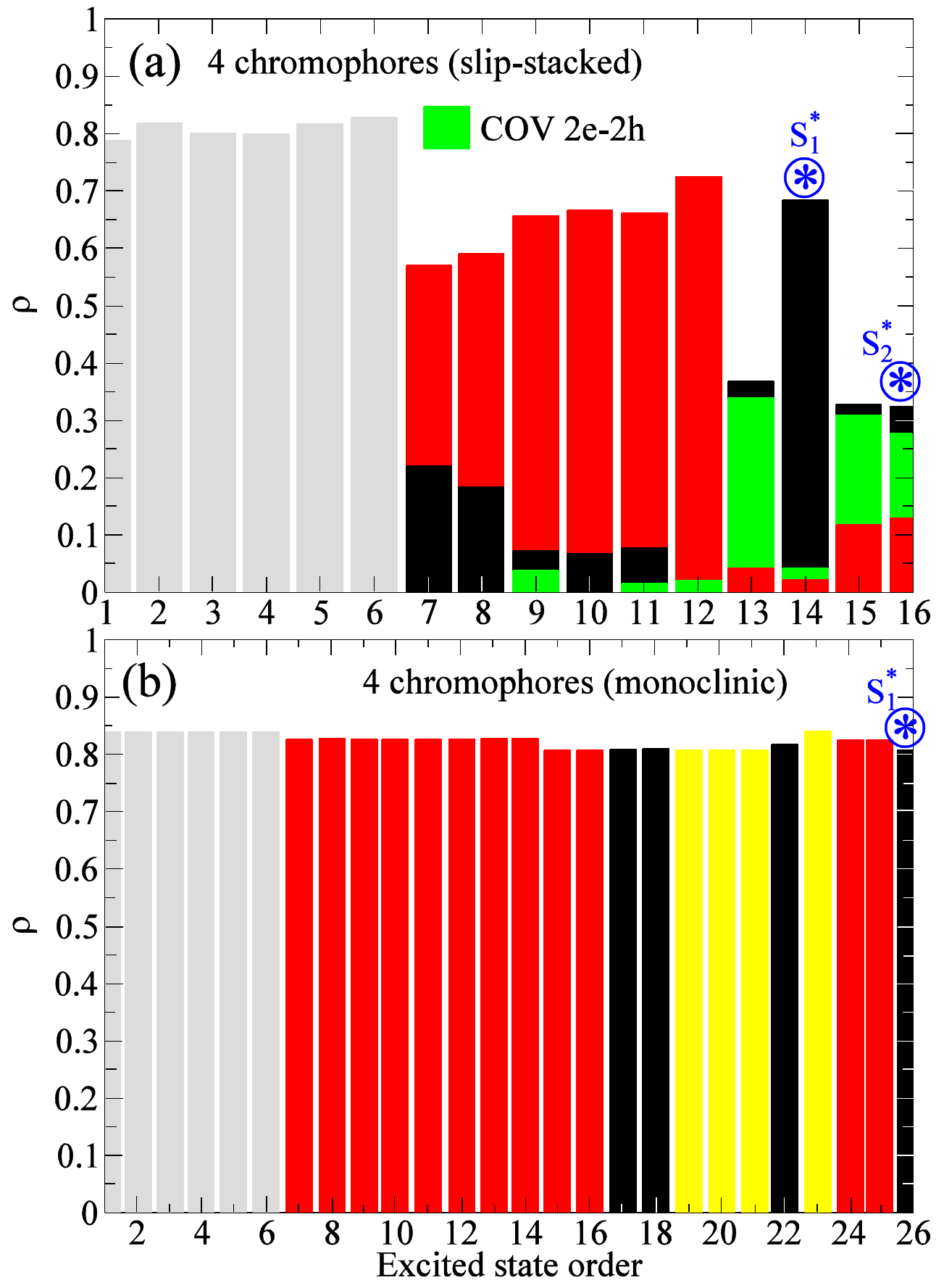} 
\protect\caption{Same as in Figure~\ref{f2}a$-$c for clusters of four octatetraene molecules, with (a) slip-stacked and (b) monoclinic geometries. For the slip-stacked geometry, eigenstates do not have $g$ and $u$ symmetry anymore, and optically allowed states are strong superpositions of 1e$-$1h and 2e$-$2h excitations whose contributions have been shown separately. The excited-state spectrum for the monoclinic lattice predicts strong SF yield, in agreement with experiments.}
\label{f3} 
\end{figure}

 {\it Slip-Stacked Geometry.} We show the results for the slip-stacked geometry for four octatetraene molecules in Figure~\ref{f3}a. Even for small intermolecular hopping, the intramolecular center of inversion is lost, and eigenstates are no longer pure $g$ or $u$. A pure one-photon forbidden 2$^1$A$_g^-$ state is thus replaced with a weakly dipole-allowed 2$^1$A$_g^-$-derived state. All optically accessible states are strong superpositions of 1e$-$1h and 2e$-$2h configurations. More importantly, the lowest eigenstates are now pure $^1$(TT) (the centers of inversion between pairs of molecules continue to persist for rigid translational shifts), and the excimer states occur above these. The slip-stacked geometry is thus conducive to SF. While our conclusion here is the same as in ref~\citenum{Smith13a}, the physical reason behind our result is very different. The previous results \cite{Smith13a} were based on symmetry characteristics of the HOMO and LUMO of pentacene. Our result is a bandwidth-induced effect that is unrelated to symmetries of the molecules or the frontier MOs; specifically, large intermolecular hopping, as would occur for the eclipsed geometry, favors the excimer over $^1$(TT). {\it Note that this conclusion emerged here only after calculations for four monomers were done and calculations based on two monomers would have predicted the incorrect result.} The smaller overall intermolecular hopping between MOs, as would occur for the slip-stacked geometry, reverses the energy ordering.
\par {\it Monoclinic Geometry.} Our calculations for the monoclinic geometry were done for four octatetraene chromophores. The wave function analysis corresponding to the states below the optical gap is presented in Figure~\ref{f3}b. Although we used the same $\beta$ for the interchromophore hopping integral (cf. eq~\ref{inter_hop}) as for the eclipsed geometry, the consequences of CT here are weaker, because of the larger intermolecular separation as well as the dependence of $t^{\perp}_{ij}$ in eq~\ref{inter_hop} on the relative orientations of the chromophores (cf. Supporting Information for details). The optical state, as well as several dark states below it, is now almost purely EXC in character. H-aggregate behavior is seen again, with the highest among these being the optically allowed excitation from the ground state. The results in Figure~\ref{f3}b are different from those for the eclipsed geometry in two major ways. First, the (TT)$_{n}$, CTX, EXC, and COV states now occur in bunches, unlike in the eclipsed case, where the energetic locations of states with the same character were more dispersed. Second, the lowest six states are now all (TT)$_{n}$ in character and are separated from the COV states (all derived from the 2$^1$A$_g^-$ in this case) by P$^+$P$^-$ excitations. The (TT)$_{n}$ states are again at energies slightly higher than $2\times E$(T$_1$) (which implies exoergic separation into triplets) and also show signatures of band formation (indicating that the triplets are not bound). Efficient SF, in agreement with experimental observation \cite{Dillon13a}, is expected here.
\par The results of Figure~\ref{f3}b are in strong support of (i) the multichromophoric nature of SF in the polyene and carotenoid molecular crystals with delocalization of the triplets to distant chromophores due to aggregation and (ii) mediation of SF by the intermediate P$^+$P$^-$ or CTX states. The mechanism of SF that one envisages from Figure~\ref{f3}b is rapid nonradiative relaxation from the optical exciton to the lowest (TT)$_n$ state, via the intermediate states. Whether or not the covalent 2$^1$A$_g^-$ states take part in the SF is difficult to predict from the energetics alone within Figure~\ref{f3}b. One possibility is that SF involves internal conversion from the 2$^1$A$_g^-$-derived states to the CTX, followed by relaxation to the (TT)$_n$. An alternate possibility involves branching of relaxation channels, whereby excitations decay to the CTX and the  2$^1$A$_g^-$ in parallel, and only the former are involved in the SF. Lower yield of triplets is expected within the second scenario. Musser et al. have argued that in the polymer P3TV, in which the 2$^{1}$A$_{g}^-$ occurs below the optical 1$^1$B$_u^{+}$, SF occurs within the single-chain but directly without going through $2^{1}$A$_{g}^{-}$ state. \cite{Musser13a} The absence of involvement of the $2^{1}$A$_{g}^{-}$ is in agreement with the picture involving branching of the excitation decay process. The \textit{single chain mechanism} however perhaps indicates bending and twisting of the long-chain polymer in the solution phase, such that different segments of the polymer are effectively decoupled. This can lead to CT between decoupled segments. We postpone the discussion of the relatively slower SF in carotenoids compared to the acenes until later. 
\par Our calculations here place the (TT)$_{n}$ states below the COV 2$^1$A$_g^-$, which we believe is necessary for efficient bimolecular SF. We have performed full CI calculations for our Coulomb interaction parameters ($U=8.0$ eV, $\kappa=2$) for octatetraene and decapentaene and have confirmed that E(2$^1$A$_g^-$)$>$ 2$\times E$(T$_1$) in the single molecules. The relative locations of the 2$^1$A$_g^-$ and the (TT)$_{n}$ depend on both the correlation strength and the physical size of the molecule, as is easily understood from physical considerations. For strong Coulomb interactions, the single triplet exciton is strongly localized, and its physical ``size'' (correlation length, which is the expectation value of the separation between the triplet-coupled spins) is small. In such a case, the 2$^1$A$_g^-$, which is a bound state of two triplet excitons, can be accommodated on a molecule that has length at least twice the size of the triplet exciton and E(2$^1$A$_g^-$)$<$ 2$\times E$(T$_1$). This is true for octatetraene and decapentaene for the standard Ohno parametrization of the PPP Hamiltonian ($U=11.26$ eV, $\kappa=1$), as was found previously from exact single-molecule calculations. \cite{Tavan87a,Ramasesha84a} We have confirmed these results also from full CI calculations. For smaller and more realistic correlations, the size of the triplet exciton is larger, and in this case the state with two spin excitations has to be ``squeezed'' to fit into the short molecules; now confinement raises the energy, making $E(2^1$A$_g^-)> 2\times E$(T$_1$). A similar argument was made previously for the bound state of two charge excitons (biexcitons) in linear chains, and numerically demonstrated. \cite{Guo95a} This effect can have an interesting consequence. On the one hand, with increasing chain length the energy of the lowest (TT)$_{n}$ state decreases relative to the 1$^1$B$_u^+$, making SF more exoergic and hence possibly more efficient. On the other hand, even with intramolecular Coulomb interactions, as the chain length is increased a length should be reached where $E(2^1$A$_g^-)$ becomes less than $2\times E$(T$_1$). Assuming that intermolecular Coulomb interactions have not raised its energy too much, from this length onward the lowest 2$^1$A$_g^-$ and the lowest (TT)$_n$ state may be very close in energy, and this interference can be detrimental to SF since a smaller fraction of the excitations will end up as $^1$(TT). In principle then, it is possible that SF efficiency reaches a peak at some intermediate molecular size. This would be interesting to examine experimentally.
\par To summarize this subsection: (i) the optical state in polyene clusters is an optically allowed CTX within the eclipsed geometry but a Frenkel exciton in the monoclinic crystal structure (several states are optically allowed with the slip-stacked geometry, but the state with the strongest oscillator strength is again a Frenkel exciton); (ii) the (TT)$_n$ states are the lowest excitations in the monoclinic structure and do not mix with the intramolecular 2e$-$2h 2$^1$A$_g^-$, indicating that SF in polyenes is multichromophoric; (iii) neither do we find eigenstates that are superpositions of the EXC and (TT)$_n$; (iv) hence the SF in the monoclinic structure is mediated by the P$^+$P$^-$ states, which are distinct eigenstates that occur between the EXC and the (TT)$_n$; and finally, (v) it is conceivable that SF yield in the polyenes peaks at some critical molecular size. We will see that the overall picture changes somewhat in the acenes.
\par
\begin{figure*}
\includegraphics[width=7.0in]{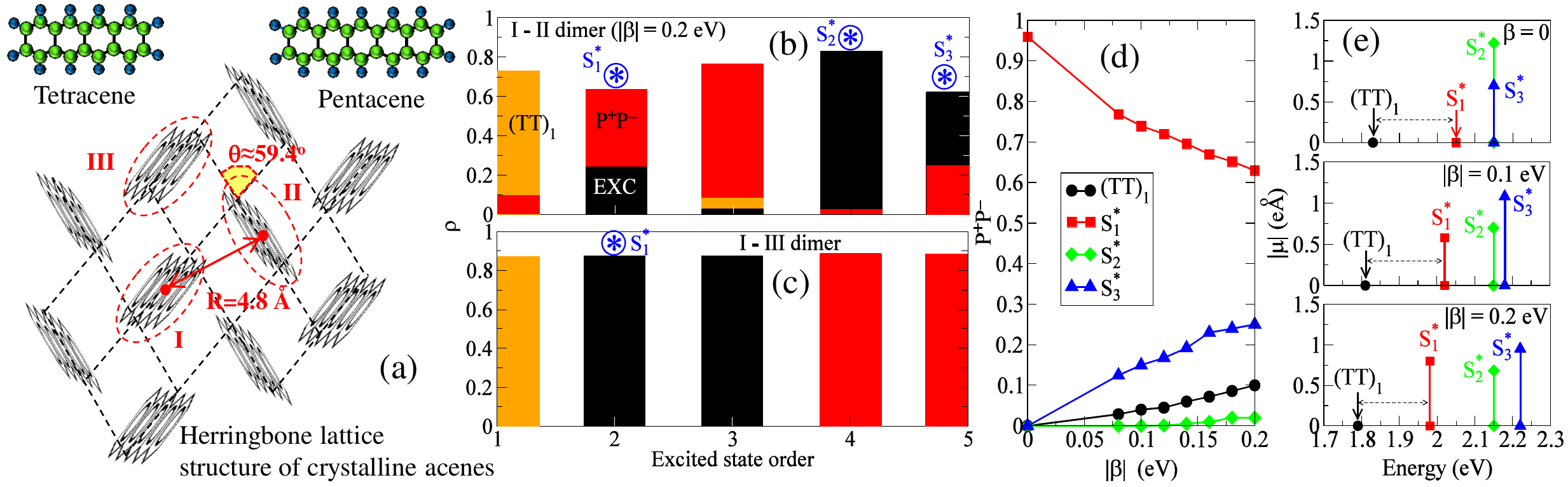} \protect\caption{(a) Herringbone lattice structure of acene crystals. Parameters $R$ and $\theta$ were taken from refs~\citenum{Kuhlman10a} and \citenum{Schiefer07a}, respectively (the parameters are known to be very close for tetracene and pentacene, and hence the same set was chosen for both). QCI calculations are performed for the I$-$II and I$-$III dimers (both pentacene and tetracene) and for the I$-$II$-$III trimer for tetracene. The wave function analysis, as in Figure~\ref{f2}, and Figure~\ref{f3} for the states proximate to the optical exciton in (b) for the I$-$II dimer and (c) for the I$-$III dimer of pentacene, respectively. (d) The relative contribution of P$^{+}$P$^{-}$ to the (TT)$_1$, S$_{1}^{\textstyle{*}}$, S$_{2}^{\textstyle{*}}$ and S$_{3}^{\textstyle{*}}$ states of the I$-$II dimer in (b) as a function of $\beta$ in eq~\ref{inter_hop}. (e) The magnitudes of the transition dipole couplings with the ground state of the allowed optical states for three different values of $|\beta|=0$, $0.1$, and $0.2$ eV. Increasing intermolecular hopping brings the lowest optical exciton closer to the (TT)$_1$ and also increases its oscillator strength.}
\label{f4} 
\end{figure*}

\noindent\textbf{Pentacene and Tetracene Crystals.} Our calculations here are for dimers (both tetracene and pentacene) and trimers (tetracene) of molecules occupying lattice sites of the characteristic herringbone crystal structure of acenes \cite{Robertson61a} shown in Figure~\ref{f4}a. We chose intramolecular Coulomb and hopping parameters slightly different from the polyenes, in order to better reproduce known single-molecule results. Thus, while for the peripheral C$-$C bonds on the circumference of the molecules we chose the standard $t_{ij}^{\mu}=-2.4$ eV, for the internal bonds that are shared by the benzene nuclei we found better agreement with experiments by using $t_{ij}^{\mu}=-2.2$ eV (the internal bonds are known to be weaker compared to the peripheral bonds \cite{Houk01a}). For the tetracene single molecule, QCI calculations with $U=5.0$ eV and $\kappa=2$ give the optical exciton and T$_{1}$ at energies $2.23$ and $1.21$ eV, compared to experimental values of $2.63$ and $1.28$ eV in solution, \cite{Murov93a} respectively ($E$(S*)$<2\times E$(T$_1$) only in crystals). For the pentacene single molecule, the experimental energies of the optical exciton and the triplet exciton are $2.12$ and $0.78$ eV, respectively.\cite{Murov93a} The corresponding quantities are $2.23$ and $0.92$ eV, respectively, from QCI calculations with $U=6.5$ eV and $\kappa=2$. In general, increasing $U$ increases the energy of the singlet optical exciton but decreases the energy of the triplet exciton; the need to fit both the singlet and the triplet exciton energies requires compromising to intermediate Coulomb strength. These reduced Coulomb interaction parameters will therefore be used for our calculations reported below for the multiple-chromophore cluster. We find the $2^{1}$A$_{g}^{-}$ at $2.57$ ($3.02$) eV for the single molecule of pentacene (tetracene), i.e., higher in energy than the optical one-photon states by $0.34$ ($0.70$) eV. Our results for the lowest two-photon state for pentacene contradicts the earlier work by Zimmerman et al., who had found an intramolecular 2e$-$2h state below the optical state \cite{Zimmerman10a} and had suggested an intrachromophore SF. Recent experimental work on SF in amorphous films \cite{Roberts12a} and solutions \cite{Walker13a} of acenes both suggest the multichromophoric nature of SF. As before, our multiple-chromophore calculations are for $\kappa^{\perp}=\kappa=2$. We have performed these calculations for a range of intermolecular $t_{ij}^{\perp}$, with $0\leqslant|\beta|\leqslant 0.2$ eV and appropriate relative angles between the $p_z$ orbitals on neighboring molecules for the herringbone crystal structure in eq~\ref{inter_hop} (cf. Supporting Information for details).
 \par {\it Pentacene.} We present our results first for pentacene since SF here is exoergic and the discussion is more straightforward. For pentacene, QCI can be implemented for a maximum of two chromophores only. In Figure~\ref{f4}b, c, we have given the same wave function analysis as for polyenes, for the dimers I$-$II and I$-$III of Figure~\ref{f4}a, respectively, for $\beta=-0.2$ eV. There is a significant difference between the two dimers, although the lowest energy state in both cases is (TT)$_1$. In the I$-$III dimer, because of the very large distances between the molecules, CT plays no role whatsoever, and higher excitations are pure EXC and P$^{+}$P$^{-}$. This is different in the I$-$II dimer, where we see three different optically allowed excitations labeled S$_1^{\textstyle{*}}$, S$_2^{\textstyle{*}}$ and S$_3^{\textstyle{*}}$, respectively, in increasing order of energy. S$_1^{\textstyle{*}}$ and S$_3^{\textstyle{*}}$ are CTX in nature, with wavefunction contributions from both EXC and P$^{+}$P$^{-}$ (S$_2^{\textstyle{*}}$ has minor P$^{+}$P$^{-}$ contribution as well). Figure~\ref{f4}d, e show the evolutions against $\beta$ of all relevant wavefunctions of the I$-$II dimer as well as their energies and transition dipole couplings with the ground state. A complete and physical picture of the consequence of CT and its role in SF is obtained from the results presented in Figure~\ref{f4}d, e, as we now discuss. At $\beta=0$, the dipole-allowed S$_2^{\textstyle{*}}$ and S$_3^{\textstyle{*}}$ are degenerate, and S$_1^{\textstyle{*}}$ is purely P$^{+}$P$^{-}$ with no transition dipole coupling to the ground state. With increasing CT ($|\beta|$), S$_1^{\textstyle{*}}$ acquires EXC character and becomes dipole-coupled to the ground state, while S$_2^{\textstyle{*}}$ loses oscillator strength as S$_3^{\textstyle{*}}$ splits from it by acquiring P$^+$P$^-$ contribution. Importantly, the (TT)$_1$ simultaneously acquires weak P$^+$P$^-$ character, and S$_1^{\textstyle{*}}$ also has moved closer to (TT)$_1$. Moreover, the (TT)$_1$ state energy $1.79$ eV $< 2\times E$(T$_1$)$=1.84$ eV, indicating the binding energy between two T$_1$ triplets due to CT ((TT)$_1$ state energy is $1.83$ eV at $\beta=0$). Taken together, one therefore sees that in pentacene crystal there occurs an optical exciton that is very close in energy to (TT)$_1$ and that both the exciton and (TT)$_1$ have contributions from P$^+$P$^-$. We believe that the proximity in energies of S$_1^{\textstyle{*}}$ and (TT)$_1$, the shared partial P$^+$P$^-$ character, and the exoergicity of the reaction lead to rapid generation of (TT)$_1$ and are behind the high SF yield in pentacene. An important difference from the polyenes therefore is that there are no intervening states here between the initial and final states. The experimentally observed Davydov splitting is 0.14 eV. \cite{Prikhotko68a} Theoretically, this is the splitting between S$_{1}^{\textstyle{*}}$ and S$_2^{\textstyle{*}}$, and from Figure~\ref{f4}e it seems $\beta=-0.16$ eV reproduces the experimental result closely. CTX character of the lowest 1e$-$1h excited states in pentacene has also been confirmed experimentally.\cite{Sharifzadeh13a,Cudazzo12a} 
\par
\begin{figure*}
\includegraphics[width=7.0in]{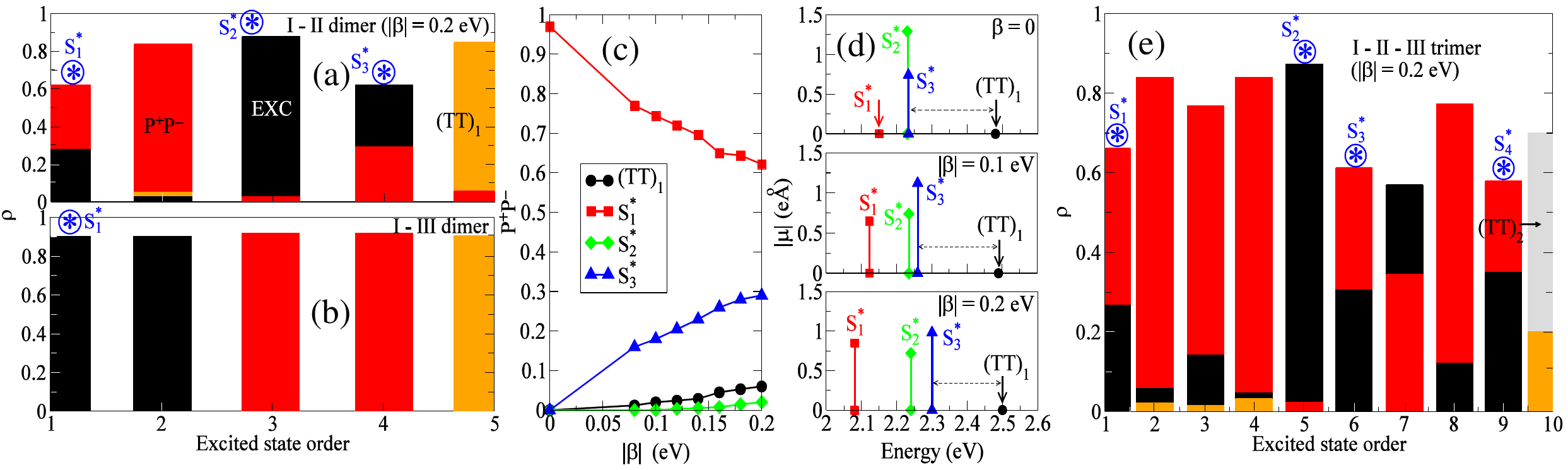} \protect\caption{Wave function analysis for the states below the lowest (TT)$_{1}$ for (a) I$-$II and (b) I$-$III tetracene dimers. (c) Same as in Figure~\ref{f4}d. (d) Same as in Figure~\ref{f4}e. (e) Wave function analysis for the I$-$II$-$III tetracene trimer. (TT)$_{1}$ and (TT)$_{2}$ correspond to I$-$II and I$-$III dimers, respectively.}
\label{f5} 
\end{figure*}

 {\it Tetracene.} For tetracene, we were able to perform QCI calculations for a cluster containing all three molecules I, II, and III in Figure~\ref{f4}a. However, as with pentacene, we first begin by looking at the I$-$II and I$-$III dimers whose energy-wave function analyses are presented in Figure~\ref{f5}a, b, again for $\beta=-0.2$ eV. Unlike pentacene, in tetracene the lowest $^1$(TT) state is energetically higher than the optical exciton, and therefore it is taken to be the highest excited state in the energy window of our analyses. For the I$-$II dimer in Figure~\ref{f5}a, the (TT)$_1$ state has energy $2.5$ eV, very close to 2$\times E$(T$_1)=2.42$ eV. Figure~\ref{f5}a, b is qualitatively similar to Figure~\ref{f4}b, c. Dimer I$-$III is again mostly irrelevant in the actual SF process, while as in pentacene in the I$-$II dimer the (TT)$_1$ state with partial P$^+$P$^-$ character is proximate to an optical exciton (S$_3^{\textstyle{*}}$ in this case) with significant P$^+$P$^-$ character. Our calculated Davydov splittings (S$_{1}^{\textstyle{*}}-$S$_2^{\textstyle{*}}$ and S$_{1}^{\textstyle{*}}-$S$_3^{\textstyle{*}}$ gaps) this time are $0.15$ and $0.22$ eV, respectively, at $\beta=-0.2$ eV, compared to the experimental $0.08$ eV, \cite{Tavazzi08a} indicating that the realistic $|\beta|$ once again is slightly smaller.
\par We show in Figure~\ref{f5}c, d the evolutions of the I$-$II tetracene dimer wave functions, energies, and transition dipole couplings to the ground states with $\beta$. The results are similar to those for pentacene. For $\beta=0$ optical excitation is to degenerate S$_2^{\textstyle{*}}$ and S$_3^{\textstyle{*}}$ excitons. With increasing $\beta$, once again the S$_1^{\textstyle{*}}$ loses P$^+$P$^-$ character, acquires EXC character, and gains oscillator strength, while the dipole-allowed S$_3^{\textstyle{*}}$ blueshifts and acquires P$^+$P$^-$ character. Simultaneously the (TT)$_1$ state also acquires weak P$^+$P$^-$ character. {\it Thus, the mechanism of SF in tetracene is very similar to that in pentacene, in that direct transition occurs from an energetically proximate optical exciton with significant P$^+$P$^-$ contribution to (TT)$_1$ with also partial P$^+$P$^-$ character.} One difference is that SF in tetracene crystals is probably still energetically uphill or at best nearly isoenergetic and, hence, less rapid than in pentacene. \cite{Smith10a,Burdett10a} In Figure~\ref{f5}e, we have given the wave function analysis for the I$-$II$-$III trimer. The new result is that now the lowest $^1$(TT) state has a contribution from the (TT)$_2$ component (triplets on I and III) that is more than three times the contribution from (TT)$_1$ (triplets on I and II), \textit{indicating a strong tendency towards triplet separation, in agreement with experimental observation}. \cite{Burdett12a} A similar tendency toward rapid triplet separation is also expected in the pentacene trimer.
\par An interesting difference between pentacene and tetracene crystals is that in the former SF is from the lowest optically allowed singlet exciton S$_1^{\textstyle{*}}$, while in the latter it is from the highest, S$_4^{\textstyle{*}}$ in Figure~\ref{f5}e. This has an interesting consequence for exciton dynamics in tetracene. Several research groups have recently pointed out that the ``prompt'' fluorescence in tetracene appears to originate from two different emitters. \cite{Burdett11a,Burdett12a,Tayebjee13a,Wilson13a} Especially at lower temperatures a fluorescence channel opens up that appears to involve an emitter different from the one involved in SF. This lower energy emitter has variously been described as a ``dark'', \cite{Burdett11a} ``dull'', \cite{Tayebjee13a} or ``trapped'' \cite{Wilson13a} exciton. Our work suggests that this is simply the S$_1^{\textstyle{*}}$ exciton and not a structural defect as suggested in ref~\citenum{Burdett11a}. Further experimental work is necessary to distinguish between the two possibilities.
\par\noindent\textbf{Covalently Linked Acene Dimers.} The high efficiency of SF in the acene crystals suggests similar high efficiency also in covalently linked dimers of such molecules. The two triplets in this case are expected to occupy the acene fragments individually. Direct covalent linking leads to too strong CT, as in bianthryl, \cite{Nishiyama98a} which we have seen is detrimental to SF (see discussion of eclipsed clusters of polyenes above). M\"uller et al. therefore investigated SF yield in three different bis(tetracene) molecules, linked through phenyl (tetracenes at para and meta positions) and biphenyl groups. \cite{Muller06a,Muller07a} In all cases the SF yield was insignificant, less than 3\% for the best case. M\"uller et al. ascribed this to the redshift of the optical exciton in the bis(tetracene) molecule, relative to tetracene; since the $^1$(TT) energy presumably stays the same in the dimers, the redshift of the optical exciton would imply larger activation energy to SF in this case. The very small shift in the optical exciton energy, as seen from the experimental solution absorption spectra (cf. Figure~1 of ref~\citenum{Muller07a}) suggests, however, that there is an additional and probably stronger reason for the tiny SF yield in the dimer (indeed, the authors have calculated the activation energies, and without assuming large differences in the prefactors in the Arrhenius rate equation, it is difficult to ascribe the gigantic difference in the SF yield to the difference in activation energy alone).
\par
\begin{figure*}
\includegraphics[width=7in]{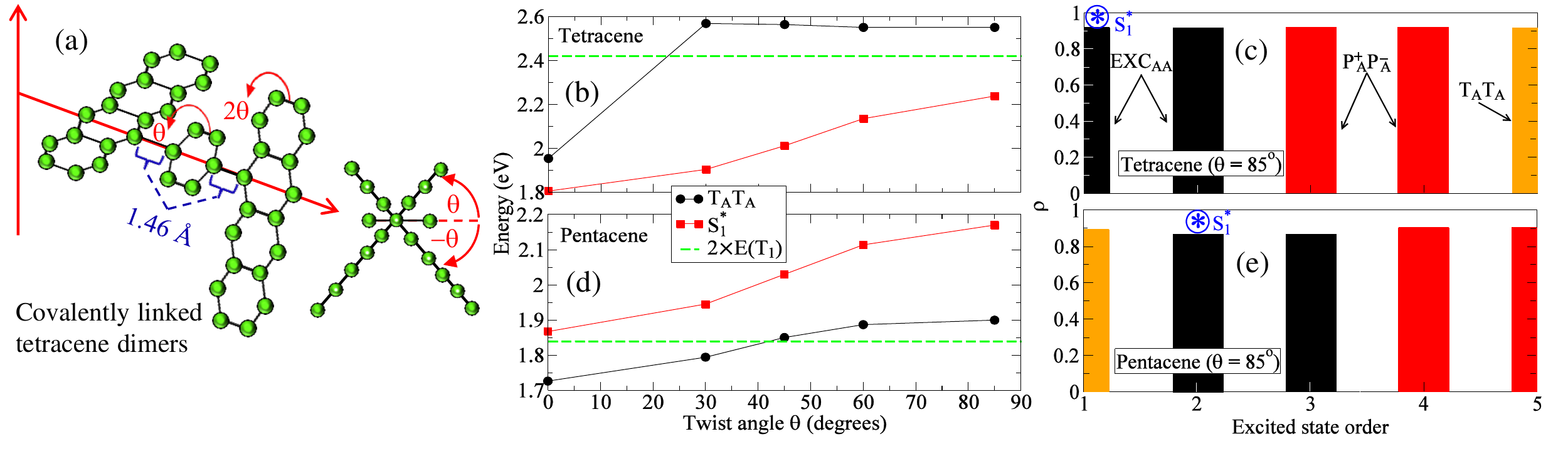} \protect\caption{(a) Two tetracene molecules linked by a phenyl group. The acene molecules are rotated by angles $\pm\theta$ with respect to the phenyl group. (b) The energies of the optical exciton and the lowest $^1$(TT) (T$_{\mathrm{A}}$T$_{\mathrm{A}}$, see text) state as a function of $\theta$ for tetracene. The horizontal dashed line is $2\times E$(T$_1$), where $E$(T$_1$) is the energy of the lowest triplet of isolated acene monomers. (c) Wave function analyses for the covalent dimer of (a), at $\theta$=85$^{\mathrm o}$ for tetracene. These wave functions are nearly the same for all $\theta > 60^{\mathrm o}$. (d) and (e) present equivalent analyses to (b) and (c), respectively, for pentacene.}
\label{f6} 
\end{figure*}

 We have theoretically investigated the bis(tetracene) molecule of Figure~\ref{f6}a, to understand the low SF yield in covalently linked dimers. Because of strong steric hindrance, the phenyl group is expected to be rotated with respect to the tetracene segments. We assume that the twist angle $\theta$ is the same in magnitude but in opposite directions for the two tetracene groups (cf. Figure~\ref{f6}a). The hopping integrals between the phenyl group and the pentacenes are taken to be $t_{\mathrm{PB}}~cos~\theta$, \cite{Ramasesha90a} where $t_{\mathrm{PB}}=-2.2$ eV is the standard hopping integral corresponding to a C$-$C single bond. We have performed QCI calculations for the bis(tetracene) molecule for varying $\theta$ and show our results for the energies of the optical exciton and the $^1$(TT) state of two triplets each located on a separate acene monomer in Figure~\ref{f6}b. Both energies rise rapidly with increasing $\theta$. The very weak redshift in the experimental solution absorption energies of the bis(tetracene) molecules indicates that the realistic $\theta$ in these molecules is large, perhaps even close to 90$^{\mathrm o}$, as is found also from {\it ab initio} calculations of the ground-state geometries. \cite{Muller07a} For such large $\theta$, we have found that excitations are almost entirely localized on the acene molecules without being delocalized also on the benzene linker. We have therefore labeled all intermolecular excitations with the suffix ``AA''. The formal expressions for EXC$_{\mathrm{AA}}$, P$_{\mathrm{A}}^+$P$_{\mathrm{A}}^-$, and T$_{\mathrm{A}}$T$_{\mathrm{A}}$ are the same as in eqs~\ref{TT-wfn} and \ref{Other-wfn}, with the only difference that the HOMO and LUMO now correspond to those of the acene units only. In Figure~\ref{f6}c, we have given the wave function analysis for the states below the lowest T$_{\mathrm{A}}$T$_{\mathrm{A}}$ at $\theta=85^{\mathrm o}$. The natures of the wave functions are now pure EXC$_{\mathrm{AA}}$, P$_{\mathrm{A}}^+$P$_{\mathrm{A}}^-$, and T$_{\mathrm{A}}$T$_{\mathrm{A}}$ as in Figure~\ref{f5}b for the I$-$III dimer of tetracene. We ascribe the low SF yield to this {\it qualitative difference in the physical nature of the wave functions}, rather than to the small difference in activation energy. Unlike in Figure~\ref{f5}a, e, where an optically allowed CTX is proximate to the (TT)$_1$ with some P$^+$P$^-$ component, states proximate to the T$_{\mathrm{A}}$T$_{\mathrm{A}}$ are pure P$^+$P$^-$ and are not optically accessible. Further, the P$^+$P$^-$ states also separate the T$_{\mathrm{A}}$T$_{\mathrm{A}}$ from the optically allowed exciton. Figure~\ref{f6}c also explains an apparently puzzling observation made by the authors of ref~\citenum{Muller07a}. The dimer according to the authors should have had two allowed exciton states, and the authors were perplexed that no such splitting of the absorption band was observed. As noted in Figure~\ref{f6}c, the second exciton state is optically dark.
\par  It is interesting in this context to also extend our calculations to the hypothetical bis(pentacene) molecule \cite{Berkelbach13b} since the issue of activation energy is not relevant in this case. We have performed the same calculations as for bis(tetracene) for bis(pentacene), assuming the same geometry. In Figure~\ref{f6}d, e, we have shown the $\theta$-dependent energies and the wave function analysis for $\theta=85^{\mathrm o}$ for this case, respectively. Once again at this large $\theta$ states are pure T$_{\mathrm{A}}$T$_{\mathrm{A}}$, EXC$_{\mathrm{AA}}$, and P$_{\mathrm{A}}^+$P$_{\mathrm{A}}^-$. Energetically, the two P$_{\mathrm{A}}^+$P$_{\mathrm{A}}^-$ states now occur even above the two EXC$_{\mathrm{AA}}$ states, only one of which is optically allowed (cf. Figure~\ref{f6}e). The wave functions are similar to those for the I$-$III dimer of pentacene in Figure~\ref{f4}c, which we have noted contributes little to SF. We predict extremely low SF here too because of weak CT, even though the reaction would be exoergic.
\section*{$\blacksquare$ CONCLUSIONS}
\label{conclusions} 
\par In summary, we have performed QCI calculations of several different classes of molecules and their clusters, with the specific goal of arriving at a comprehensive physical picture of SF in these systems. The attraction of the semiempirical approach is that it allows the inclusion of a very large number of basis functions, making it possible to treat all 1e$-$1h and 2e$-$2h excited states on equal footing. This is necessary for arriving at the correct energy ordering of the excited states in the clusters, which we have argued is essential to understanding SF. The widely recognized inequality $E$(S*$)\gtrsim 2\times E$(T$_1$) is a necessary but not sufficient condition for high SF efficiency. On the other hand, the PPP model QCI calculations allow us to explain the similarities as well as differences in SF between three different classes of materials without making additional assumptions. 
\par For all the cases we have examined, we find that intermolecular CT plays a strong role in SF. The precise role and extent of CT, are, however, strongly dependent on structure and morphology. In the acenes SF involves a one-step transition from a proximate singlet optical exciton to the $^1$(TT) state, where both the initial and final states share P$^+$P$^-$ character. The very high yields and rate of SF in pentacene are consequences of the unique electronic structures of both the optical exciton and the $^1$(TT) states in this case. Our proposed mechanism here is very close to that given by Beljonne et al., \cite{Beljonne13a} for equilibrium separations between the pentacene molecules. We do not find a coherent superposition of EXC, P$^+$P$^-$, and $^1$(TT) with significant contributions from all three, and indeed such a superposition is unexpected in the absence of direct coupling between EXC and $^1$(TT). Such a superposition has been proposed by Chan et al. \cite{Chan11a} and has also been calculated by Beljonne et al., for intermolecular separations smaller than equilibrium separations, when the large first order couplings between EXC-P$^+$P$^-$ and $^{1}$(TT)-P$^+$P$^-$ pairs can give large second-order coupling between EXC and $^1$(TT) if the P$^+$P$^-$ state energy is not too large. Similar results were also found by Zeng et al., \cite{Zeng14a} who had calculated the eigenstates of the pentacene I-II dimer in the reduced space of the six basis functions that are obtained when only the HOMO and LUMO of each molecule are retained in the calculation (the ground state with both LUMOs unoccupied, two EXC and P$^+$P$^-$, and a single TT configuration as the basis components). Zeng et al.'s eigenstates are qualitatively similar to ours for equilibrium geometry, except that they also find a state that is a strong superposition of EXC, P$^+$P$^-$, and $^1$(TT) when the couplings between the molecules are made artificially large by either bringing them closer to one another or by rotating them. In neither of these cases the authors calculate the Davydov splittings for the increased intermolecular couplings. It is our belief that the resultant  Davydov splittings would be far larger than the experimentally observed quantities. This is substantiated by our determination that while with $|\beta|\sim0.16$ eV in eq~\ref{inter_hop} we are able to reproduce the experimental Davydov splitting of $\sim0.14$ eV for pentacene even for a modest increase to $|\beta|=0.2$ eV in eq~\ref{inter_hop} calculated splittings exceed the observed values. We further believe that the mixing between EXC and $^1$(TT) is reduced further for calculations within the complete basis space (as opposed to the reduced space of six basis functions) since all three classes of eigenstates,  EXC, P$^+$P$^-$, and $^1$(TT), now have contributions from higher energy excitations, CT involving which is less efficient. Corresponding to each of our normalized many-body wave function calculated with a complete to nearly complete basis, it is possible to calculate the error in the computed wave function had only the six basis functions of Zeng et al. \cite{Zeng14a} been retained. This is simply the total sum of the contribution to the wave function by all other basis functions that are ignored within the reduced basis space. Proceeding in this manner we have determined that while the errors in the calculations of EXC and CTX in pentacene are 16\% and 12\%, respectively, the corresponding error for $^1$(TT) is 20\%. 
\par In contrast to the acenes, in polyenes with the monoclinic crystal structure, the transition is from a pure optical exciton to a pure $^1$(TT) and is at least a two-step process, via P$^+$P$^-$ states. Whether it is even a three-step process with the 2$^1$A$_g^-$ as yet another intermediate or whether the relaxation of the singlet optical exciton occurs via two competing paths, one involving the 2$^1$A$_g^-$ and the other leading to SF, cannot be predicted from our calculations. In either case we expect slower generation rate of triplets through SF in the polyenes than in pentacene, in agreement with observations. \cite{Gradinaru01a,Papagiannakis02a,Wang10a,Wang12a,Antognazza10a} Finally, errors incurred if the calculations of $^1$(TT) states in polyenes are done in the reduced space of six basis functions are huge, close to 50\%. This is because bonding (antibonding) MOs much lower (higher) than the HOMO (LUMO) make very significant contributions to the $^1$(TT) here.
\par The roles of CT are extreme in the polyenes with the hypothetical eclipsed geometry and the covalently linked acene dimers with realistic geometry. In the former, CT is too strong, and as a consequence the lowest excited state is an excimer, rather than $^1$(TT). Efficient SF is not expected here. There are many examples of real systems where excimer formation precludes SF. \cite{Nishiyama98a} The slip-stacked geometry is more favorable for SF than the eclipsed geometry because of weaker CT. In the covalently linked acene dimers, in contrast, there is too little CT. The $^1$(TT) states in bis(tetracene) are separated from the lower energy optical exciton by the optically dark P$^+$P$^-$ states at intermediate energy, precluding transition from the optical exciton to the $^1$(TT). This explains the very low SF yield here. We predict similar low SF yield also in the hypothetical bis(pentacene). 
\par The calculations we have reported are for a purely electronic Hamiltonian, and vibronic couplings have not been included. The effects of including vibronic couplings can, however, be guessed based on the general observation that vibronic couplings lower the energies of molecular ionic states more than that for neutral molecules. The same principle applies to excited states, where vibronic couplings lower the energies of charge-separated ionic states more than charge-neutral covalent states. Furthermore, the larger the electron$-$hole separation, the greater the energy lowering. Thus, energies of polaron pairs and charge-transfer exciton states, with the charges on separate molecules, will be lowered more than those of the intramolecular exciton, and the latter's energy will be lowered more than triplet$-$triplet states. From Figure~\ref{f4}b, this would imply even larger SF rate for pentacene where S$_1^{\textstyle{*}}$ becomes more proximate to $^1$(TT). Conversely, S$_4^{\textstyle{*}}$ in Figure~\ref{f5}e becomes less proximate to the $^1$(TT) above it in tetracene, and the SF rate is reduced slightly. In the monoclinic carotenoids, given the very large numbers of states involved not much change is expected. In the case of the bis(tetracene), vibronic couplings will not make any difference, as the states closest to the triplet$-$triplet are pure polaron pairs that are not optically accessible.
\par Finally, our work here has focused on the mechanism of SF alone and not whether the molecules we have investigated will be particularly useful in enhancing photovoltaic performance. A necessary condition for the triplets undergoing charge dissociation at a donor$-$acceptor interface is that the donor$-$acceptor exciplex has even lower energy. \cite{Aryanpour13a} This condition is currently satisfied by relatively few heterostructures, the most well-known example of which is the pentacene$-$C$_{60}$ solar cell. The low triplet energy in this case is the reason behind the very small open-circuit voltage, \cite{Lee09a,Congreve13a,Jadhav12a} which reduces the performance of the solar cell even though the number of charge carriers is large. Organic donor molecules with significantly larger triplet energy than acenes are therefore of interest, and we are aware of experimental investigations of polycyclic aromatic hydrocarbons (PAHs) with large optical and triplet gaps as candidates for efficient SF. \cite{Nichols13a} We have recently performed theoretical work on excited state orderings in large PAH molecules. \cite{Aryanpour14b} The approach we have developed here can in principle be extended to clusters of such PAH molecules. This is the goal of future work.
\section*{$\blacksquare$ ASSOCIATED CONTENT}
\noindent \textbf{Supporting Information.} Details of QCI calculations, intermolecular hopping integrals between octatetraene and acene chromophores in the monoclinic and herringbone lattice structures, respectively, and the 2$^1$A$_g^-$ state energy in the eclipsed, slip-stacked, and monoclinic arrangements of hexatriene chromophores. This material is available free of charge via the Internet at http://pubs.acs.org.
\vskip 1pc
\section*{$\blacksquare$ AUTHOR INFORMATION}
\noindent \textbf{Corresponding Author} \\
*Phone: +1-520-621-6798. Fax: +1-520-621-4721. E-mail: karana@physics.arizona.edu \\
 \textbf{Notes} \\
 The authors declare no competing financial interest. 
\section*{$\blacksquare$ ACKNOWLEDGMENTS}
Work at Arizona was supported by NSF grant CHE-1151475. A travel grant awarded by the Indo-US Science and Technology Forum, Award Number 37-2012/2013-14, facilitated the international collaboration.
\section*{$\blacksquare$ NOTE ADDED IN PROOF}
\noindent After this manuscript was accepted for publication, we were notified about the new publication by T. C. Berkelbach et al.,  \cite{Berkelbach14a} in which the authors have also discussed CT character of the low-lying excited states in crystalline pentacene.
\clearpage{}\newpage{}\onecolumngrid \pagenumbering{arabic}
\pagestyle{fancy}
\fancyhf{}
\renewcommand{\headrulewidth}{0pt}
\cfoot{S\thepage}
\begin{center}
\textbf{\large SUPPORTING INFORMATION: Theory of Singlet Fission in Polyenes, Acene Crystals and Covalently Linked Acene Dimers}
{\large\vskip 1pc Karan Aryanpour$^1$, Alok Shukla$^2$, and Sumit Mazumdar$^{1,3}$} 
\vskip 1pc $^1$Department of Physics, University of Arizona, Tucson, AZ 85721, United States \\
$^2$Department of Physics, Indian Institute of Technology Bombay, Powai, Mumbai - 400076, India \\
$^3$College of Optical Sciences, University of Arizona, Tucson, AZ 85721, United States
\end{center}
\textbf{\large Contents:}
\begin{enumerate}
\item Details of QCI calculations  
\item Monoclinic lattice structure and intermolecular hopping integrals for four octatetraene chromophores 
\item Intermolecular hopping integrals for acene chromophores in herringbone lattice structure 
\item 2$^1$A$_g^-$ state energy in the eclipsed, slip-stacked and monoclinic arrangements of hexatriene chromophores
\end{enumerate}
\vskip 1pc 
\textbf{\large 1. Details of QCI calculations} \\
\par Number of C atoms, active MOs and dimensions of the final QCI matrices
for octatetraene (cf. Figure~1a), decapentaene, pentacene,
tetracene (cf. Figure~4a) and covalently linked dimers (cov-dim) of tetracene (cf. Figure~6a) and pentacene multiple chromophores are all given
in the following table: 
\\
\\
\begin{tabular}{l|l|l|l|l|l}
\hline 
{}&{\footnotesize ~~~~~~~~~octatetraene:}&{\footnotesize ~~~decapentaene:}&{\footnotesize pentacene:}&{\footnotesize ~~~~~tetracene:}&{\footnotesize tetracene (pentacene) cov-dim:}\tabularnewline
{\footnotesize Chromophore number:}&{\footnotesize ~~~2\hspace{1.15cm}3\hspace{1.15cm}4}&{\footnotesize ~~~~2\hspace{1.2cm}3}&{\footnotesize ~~~~~~2}&{\footnotesize ~~~~~2\hspace{1.15cm}3}&{\footnotesize ~~~~~~~~~~~~~~~~~~~~3}\tabularnewline
{\footnotesize C atom number:}&{\footnotesize ~~16\hspace{1cm}24\hspace{1cm}32}&{\footnotesize ~~~20\hspace{1.1cm}30}&{\footnotesize ~~~~~44}&{\footnotesize ~~~~36\hspace{1cm}54}&{\footnotesize ~~~~~~~~~~~~~~~~~~~42 (50)}\tabularnewline
{\footnotesize Active MO number:}&{\footnotesize ~~16\hspace{1cm}18\hspace{1cm}16}&{\footnotesize ~~~20\hspace{1.1cm}18}&{\footnotesize ~~~~~20}&{\footnotesize ~~~~20\hspace{1cm}18}&{\footnotesize ~~~~~~~~~~~~~~~~~~~20 }\tabularnewline
{\footnotesize Matrix size:}&{\footnotesize$294185$\hspace{0.4cm}$772435$\hspace{0.4cm}$294185$}&{\footnotesize$1833276$\hspace{0.4cm}$772435$}&{\footnotesize ~~$1833276$}&{\footnotesize $1833276$\hspace{0.4cm}$772435$}&{\footnotesize ~~~~~~~~~~~~~~~~$1833276$ }\tabularnewline
\hline 
\end{tabular}
\\
\\
{\footnotesize Table~S1: Details of QCI calculations for multiple chromophores of polyenes and acenes studied in the main article.}
\\
\\
\\
{\textbf{\large 2. Monoclinic lattice structure and intermolecular hopping integrals for four octatetraene chromophores} }\\
\par According to refs~[S1, S2], diphenylpolyenes
usually crystallize in monoclinic and orthorhombic lattice structures.
The monoclinic structure was adopted for this work in which more efficient
SF compared to the orthorhombic structure was reported. [S3]
We take an octatetraene chromophore originally lying in the \textit{xy}
plane along the \textit{x} axis with its center on the origin. We
rotate this chromophore around the \textit{x} axis by $\gamma=\pm30^{\circ}$
followed by another rotation around the \textit{z} axis by $\theta=42.30^{\circ}$,
resulting in two new chromophores. [S1] A copy of the
one rotated by $\gamma=+30^{\circ}$ is translated by distance $a=6.23\mathring{\textrm{A}}$
along the \textit{x} axis while the other rotated by $\gamma=-30^{\circ}$
and a copy of it by distances $\frac{1}{2}a$ along the \textit{x}
and $\pm\frac{1}{2}b$ along the \textit{z} axes with $b=7.45\mathring{\textrm{A}}$
(cf. Figure~1c). [S2] Using eq~6
in which $\beta=-0.2$ eV and $h_{min}=3.5\mathring{\textrm{A}}$
as chosen for the eclipsed geometry (cf. Figure~1a) and also considering the angles between the $p_z$ orbitals and the lines connecting them on two separate chromophores, the largest calculated $t^{\perp}$ value was $-0.04$ eV.\\
 \\
\begin{figure*}
\includegraphics[width=4.5in]{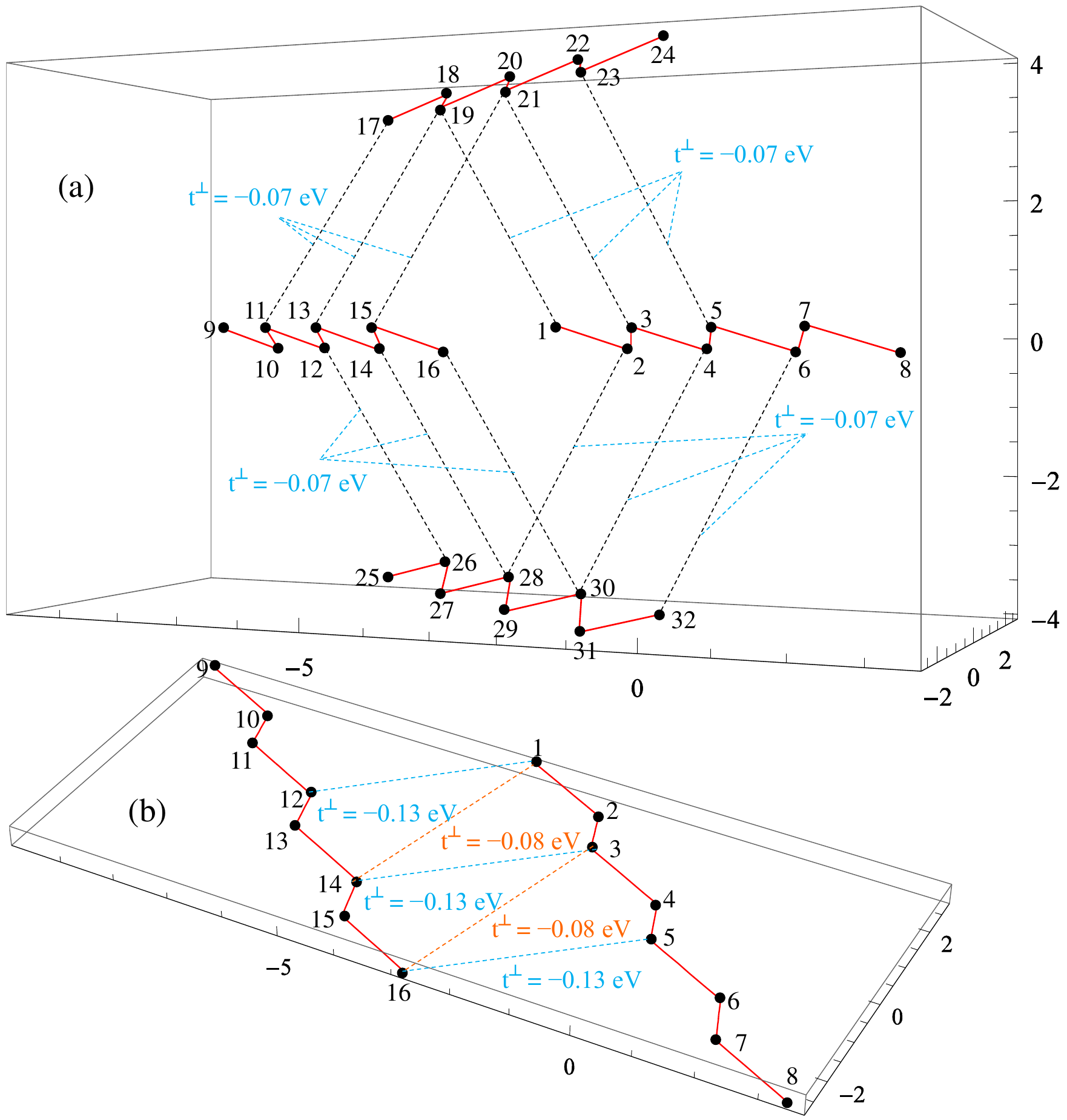} \vspace{0.2cm}
\footnotetext{\hspace{-0.3cm}{\small Figure~S1: (a) Side and (b)
top views of intermolecular hopping integrals
among four octatetraene chromophores in the monoclinic lattice structure
with lattice parameters given in Figure~1c. Angles $\theta_{1-4}$ represent the relative orientations of the $p_{z}$ orbitals on two separate chromophores.}
 \label{S1} }
\end{figure*}
\\
{\textbf{\large 3. Intermolecular hopping integrals for acene chromophores in the herringbone lattice structure} }\\
\par For acene chromophores possessing herringbone lattice structure
with parameters given in Figure~4a, we employed eq~6
with $\beta=-0.2$ to $-0.08$ eV in steps of $0.02$ eV (cf. Figure~4d, 5c) and $h_{min}=3.50\mathring{\textrm{A}}$ to
compute $t^{\perp}$ values. Taking account of the relative angles between the $p_{z}$ orbitals, only two intermolecular hopping integrals, i.e., $t^{\perp}_{1}$ and $t^{\perp}_{2}$, become greater than $0.07$ eV, our intermolecular hopping integral threshold, for the largest $\beta=-0.2$ eV as shown in Figure~S2 for two pentacene chromophores. Our chosen range of $\beta$ is consistent
with the weak to intermediate CT effects between I$-$II and II$-$III chromophores
in Figure~4a in comparison with experiments. [S4, S5]
Within this range, our calculated $t^{\perp}$ values as presented below between the sites
numbered for two pentacene chromophores depicted in Figure~S2: \\
 \\
 $\beta=-0.2$ eV:\\
 $t^{\perp}_{1}=-0.13$ eV: $(24,5)$, $(30,9)$, $(34,13)$, $(38,17)$, $(42,21)$.
\\
 $t^{\perp}_{2}=-0.07$ eV: $(24,4)$, $(24,6)$\hspace{0.2in}$(30,6)$, $(30,10)$\hspace{0.2in}$(34,10)$,
$(34,14)$\hspace{0.2in}$(38,14)$, $(38,18)$\hspace{0.2in}$(42,18)$,
$(42,22)$. \\
\\ \\
 and \\ \\ 
$\beta=-0.18$ eV:\\
 $t^{\perp}_{1}=-0.12$ eV, $t^{\perp}_{2}=-0.07$ eV \\ \\
$\beta=-0.16$ eV:\\
$t^{\perp}_{1}=-0.11$ eV, $t^{\perp}_{2}\approx 0$ \\ \\
$\beta=-0.14$ eV:\\
$t^{\perp}_{1}=-0.09$ eV, $t^{\perp}_{2}\approx 0$ \\ \\
$\beta=-0.12$ eV:\\
$t^{\perp}_{1}=-0.08$ eV, $t^{\perp}_{2}\approx 0$ \\ \\
$\beta=-0.1$ eV:\\
$t^{\perp}_{1}=-0.07$ eV, $t^{\perp}_{2}\approx 0$ \\ \\
$\beta=-0.08$ eV:\\
$t^{\perp}_{1}=-0.06$ eV, $t^{\perp}_{2}\approx 0$ \\ \\
where $|t^{\perp}|<0.07~\textrm{eV}\approx0$. The same $t^{\perp}$ applies to the equivalent sites of two tetracene
chromophores. 
\begin{figure*}
\includegraphics[width=4.5in]{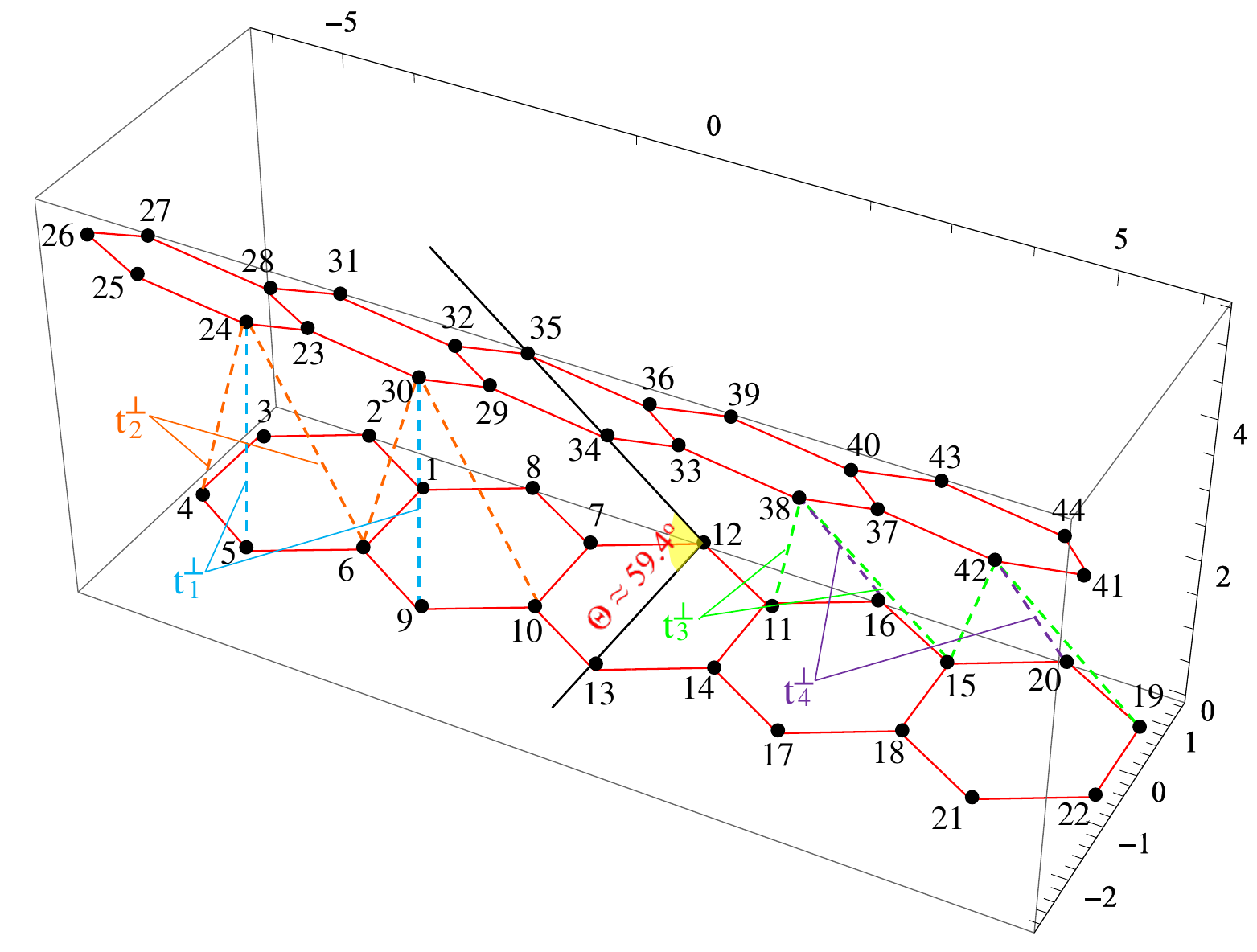} \vspace{0.2cm}
\footnotetext{\hspace{-0.3cm}{\small Figure~S2: View of intermolecular hopping integrals between two pentacene chromophores in the herringbone
lattice structure with lattice parameters given in Figure~4a. Angles $\theta_{1-4}$ represent the relative orientations of the $p_{z}$ orbitals on two separate chromophores.}
\label{S2} }
\end{figure*}
\\
\\
{\textbf{\large 4. 2$^1$A$_g^-$ state energy in the eclipsed, slip-stacked and monoclinic arrangements of hexatriene chromophores}}\\ 
\par QCI results for the 2$^1$A$_g^-$ state energy of hexatriene chromophores are presented below as they are added one by one in the eclipsed geometry (cf. Figure~1a) and also when four chromophores are arranged together in the slip-stacked and monoclinic geometries (cf. Figure~1b, c). The number of MOs included in QCI calculations is also specified (energies are in eV): \\
 \\ \\
\begin{tabular}{l}
\hline 
{\footnotesize Number of chromophores:\hspace{1.6cm}$1$\hspace{1.7cm}$2$\hspace{2.2cm}$3$\hspace{2.1cm}$4$\hspace{2.3cm}$4$\hspace{2.5cm}$4$}\tabularnewline
{\footnotesize {Number of MOs included in QCI:}\hspace{0.6cm}$6$\hspace{1.0cm}(eclipsed) $12$\hspace{0.6cm}(eclipsed) $18$\hspace{0.6cm}(eclipsed) $16$\hspace{0.6cm}(slip-stacked) $16$\hspace{0.6cm}(monoclinic) $16$}
\tabularnewline
{\footnotesize\hspace{4.4cm}$E_1 = 3.98$\hspace{0.6cm}$E_1 = 4.13$\hspace{1.0cm}$E_1 = 4.45$\hspace{0.9cm}$E_1 = 4.76$\hspace{1.2cm}$E_1 = 4.67$\hspace{1.5cm}$E_1 = 4.93$}\tabularnewline
{\footnotesize\hspace{6.3cm}$E_2 = 4.16$\hspace{1.0cm}$E_2 = 4.51$\hspace{0.9cm}$E_2 = 4.81$\hspace{1.2cm}$E_2 = 4.70$\hspace{1.5cm}$E_2 = 4.93$}\tabularnewline
{\footnotesize\hspace{8.55cm}$E_3 = 4.52$\hspace{0.95cm}$E_3 = 4.87$\hspace{1.2cm}$E_3 = 5.12$\hspace{1.5cm}$E_3 = 4.93$}\tabularnewline
{\footnotesize\hspace{10.8cm}$E_4 = 4.98$\hspace{1.2cm}$E_4 = 5.16$\hspace{1.5cm}$E_4 = 5.04$}\tabularnewline
\hline 
\end{tabular} \\
\\
{\footnotesize Table~S2: QCI results for the energy of 2$^1$A$_g^-$ states in the eclipsed, slip-stacked and monoclinic arrangements of hexatriene chromophores.}
\\
\\
\\
As it is readily seen, 2$^1$A$_g^-$ state energy increases with the number of chromophores. \\
\\
\\
{\textbf{\large References}}
\\
\\
S1.~Drenth, W.; Wiebenga, E. H. Structure of $\alpha$.$\omega$-Diphenylpolyenes: II. Crystal Structure of the Monoclinic and Orthorhombic Modification of 1,10-Diphenyl-1,3,5,7,9-decapentaene. \textit{Recl Trav.\ Chim.\ Pays-Bas.}\textbf{1954,} \textit{73,} 218$-$228. 
\\ \\
S2.~Weiss, V.; Port, H.; Wolf, H. C. Excitonic and Molecular Properties of the Triplet T$_1$ State in Diphenylpolyene Single Crystals. \textit{Mol.\ Cryst.\ Liq.\ Cryst.} \textbf{1997,} \textit{308,} 147$-$178. 
\\ \\
S3.~Dillon, R. J.; Piland, G. B.; Bardeen, C. J. Different Rates of Singlet Fission in Monoclinic versus Orthorhombic Crystal Forms of Diphenylhexatriene. \textit{J.\ Am.\ Chem.\ Soc.} \textbf{2013,} \textit{135,} 17278$-$17281.
\\ \\
S4.~Prikhotko, A. F.; Tsikora, L. I. Spectral Investigations of Pentacene. \textit{Opt. Spectrosc.} \textbf{1968,} \textit{25,} 242$-$246. 
\\ \\
S5.~Tavazzi, S.; Raimondo, L.; Silvestri, L.; Spearman, P.; Camposeo, A.; Polo, M.; Pisignano, D. Dielectric Tensor of Tetracene Single Crystals: The Effect of Anisotropy on Polarized Absorption and Emission Spectra. \textit{ J.\ Chem.\ Phys.} \textbf{2008,} \textit{128,} 154709.

\end{document}